\documentclass[12pt]{article}
\usepackage{amsmath}
\usepackage{amssymb}
\usepackage{amsthm}
\usepackage{fullpage}
\usepackage[utf8]{inputenc}
\usepackage{mathtools}
\usepackage{url}
\usepackage[usenames,svgnames]{xcolor}
\usepackage{cite}

\usepackage[pagebackref=false]{hyperref} 
\usepackage[all]{hypcap} 

\theoremstyle{plain}
\newtheorem{theorem}{Theorem}[section]
\newtheorem{lemma}[theorem]{Lemma}

\newtheorem{proposition}[theorem]{Proposition}
\newtheorem{corollary}[theorem]{Corollary}
\newtheorem{example}{Example}[section]
\newtheorem{examples}{Example}[subsection]

\newtheorem{remark}{Remark}[section]

\theoremstyle{definition}
\newtheorem{definition}{Definition}[section]

\numberwithin{equation}{section} 

\hypersetup{
breaklinks,
colorlinks,
citecolor=Green,
linkcolor=BlueViolet,
urlcolor=DarkBlue,
pdftitle={Quantum Hurwitz numbers and MacDonald polynomials},
pdfauthor={John Harnad }, }

\DeclareMathOperator{\tr}{tr}
\DeclareMathOperator{\End}{End}
\DeclareMathOperator{\Span}{span}

\DeclareMathOperator{\cyc}{cyc}

\DeclareMathOperator{\ch}{ch}

\DeclareMathOperator{\Li}{Li}
\DeclareMathOperator{\aut}{aut}

\def\ra{{\rightarrow}}

\def\mt{{\mapsto}}

\def\tr{\mathrm {tr}}
\def\det{\mathrm {det}}

\def\End{\mathrm {End}}
\def\span{\mathrm {span}}

\def\ln{\mathrm {ln}}

\def\span{\mathrm {span}}

\DeclarePairedDelimiter{\abs}{|}{|}

\DeclarePairedDelimiter{\ket}{|}{\rangle}

\def\be{\begin{equation}}
\def\ee{\end{equation}}

\def\bea{\begin{eqnarray}}
\def\eea{\end{eqnarray}}

\def\bt{\begin{theorem}}
\def\et{\end{theorem}}

\def\bex{\begin{example}\small \rm}
\def\eex{\end{example}}

\def\bexs{\begin{examples}\small \rm}
\def\eexs{\end{examples}}

\def\ra{\rightarrow}
\def\ss{\subset}
\def\deq{\coloneqq}

\def\br{\begin{remark}\small \rm}
\def\er{\end{remark}}

\def\mt{{\mapsto}}

\def\&{&{\hskip -20pt}}

\def\BB{\mathcal{B}}

\def\EE{\mathcal{E}}
\def\FF{\mathcal{F}}
\def\JJ{\mathcal{J}}
\def\HH{\mathcal{H}}
\def\II{\mathcal{I}}

\def\OO{\mathcal{O}}
\def\PP{\mathcal{P}}

\def\ZZ{\mathcal{Z}}

\def\Bb{\mathbf{B}}
\def\Cb{\mathbf{C}}
\def\Ib{\mathbf{I}}

\def\Nb{\mathbf{N}}
\def\Pb{\mathbf{P}}

\def\Zb{\mathbf{Z}}

\def\Zbb{\mathbb{Z}}

\def\Zbb{\mathbb{Z}}

\def\grG{\mathfrak{G}} \def\grg{\mathfrak{g}}

 \def\grl{\mathfrak{l}}

 \def\grgl{\mathfrak{gl}}

\begin{document}
\baselineskip 16pt
\medskip
\begin{center}
\begin{Large}\fontfamily{cmss}
\fontsize{17pt}{27pt}
\selectfont
\textbf{Weighted Hurwitz numbers and hypergeometric $\tau$-functions: an overview}\footnote{Work  supported by the Natural Sciences and Engineering Research Council of Canada (NSERC) and the Fonds Qu\'ebecois de la recherche sur la nature et les technologies (FQRNT). }
\end{Large}\\
\bigskip
\begin{large}  {J. Harnad}$^{1,2}$
 \end{large}
\\
\bigskip
\begin{small}
$^{1}${\em Centre de recherches math\'ematiques,
Universit\'e de Montr\'eal\\ C.~P.~6128, succ. centre ville, Montr\'eal,
QC, Canada H3C 3J7 } \\
\smallskip
$^{2}${\em Department of Mathematics and
Statistics, Concordia University\\ 1455 de Maisonneuve Blvd.~W.  
Montr\'eal, QC,  Canada H3G 1M8 } 
\end{small}
\end{center}
\bigskip

\begin{abstract}
   This  is  a survey of  recent developments in the use of  2D Toda $\tau$-functions of  hypergeometric type as  generating functions for  multiparametric families of weighted Hurwitz numbers.  Such $\tau$-functions are obtained by using diagonal group elements in their expression as fermonic vacuum expectation values, implying that their expansion in a basis of tensor products of Schur functions is diagonal. A corresponding abelian group action on the center of the $S_n$ group algebra is defined by evaluating symmetric functions formed multiplicatively from a weight generating function $G(z)$ on the Jucys-Murphy elements of the group algebra.  The resulting central elements act diagonally in the basis  of orthogonal idempotents,  with eigenvalues  $r^{G(z)}_\lambda$ that coincide with the coefficients  in the double Schur function expansion. The group action is represented  in the basis  of  cycle sums by matrices whose  elements, expanded as power series in $z$,  are the weighted double Hurwitz numbers. Both their geometrical meaning, as weighted sums over $n$-sheeted branched coverings, and  combinatorial one,  as weighted enumeration of paths in the Cayley graph of $S_n$  generated by transpositions,  follow from expanding the Cauchy-Littlewood generating functions over dual pairs of bases of the algebra of symmetric functions and evaluating on the Jucys-Murphy elements. It follows that the coefficients in the  expansion of  the $\tau$-function  in the basis of products of  power sum symmetric functions are the weighted Hurwitz numbers. All previously studied cases are obtained by making suitable  choices for  $G(z)$. Expansion in powers of some  of the parameters  determining the weighting  provide generating series for multispecies weighted Hurwitz numbers.  Replacement of the Cauchy-Littlewood  generating function by the one  for Macdonald polynomials provides $(q,t)$-deformations that are generating functions for quantum weighted Hurwitz numbers.  
     \end{abstract}

\break
\tableofcontents

\section{Hurwitz numbers }

The study of Hurwitz numbers, which enumerate branched covers of the Riemann sphere with specified ramification profiles, began with the pioneering work of Hurwitz \cite{Hu1, Hu2}. Their relation to enumerative factorization  problems in the symmetric group and  irreducible characters was developed by  Frobenius \cite{Frob1, Frob2} and Schur \cite{Sch}.  In recent years,  following the discovery by Pandharipande \cite{Pa} and Okounkov \cite{Ok} that certain KP and  2D Toda $\tau$-functions \cite{Ta, UTa, Takeb}, fundamental to the modern theory of integrable systems \cite{Sa, SS},  could serve as  generating functions for weighted Hurwitz numbers,  there has been a flurry of activity \cite{GGN1, GGN2, GH1, GH2, AMMN, HO3, AC1, AC2, BEMS, Z, KZ, H1, NOr1, NOr2, H2} concerned with  finding new classes of $\tau$-functions  that can similarly  serve as generating functions for various types of weighted Hurwitz numbers.
 
Two closely related interpretations of these weighted Hurwitz numbers exist. The enumerative geometrical one consists of weighted sums of Hurwitz numbers for $n$-sheeted branched coverings of the Riemann sphere. The other consists of weighted enumeration of factorizations of elements of the symmetric group $S_n$  in which the factors are in specified  conjugacy classes. This may equivalently be interpreted as  a weighted counting  of paths in the Cayley graph generated by  transpositions, starting and ending in specified classes. The  two  approaches are related by the monodromy representation of the fundamental group of the sphere   punctured at the branch points  obtained by lifting closed paths to the covering surface. Variants of this also exist for branched coverings of higher genus surfaces \cite{LZ}  and other groups.

Some  generating functions of enumerative invariants are known to also have representations as matrix integrals \cite{GGN1, GGN2, GH1, AC1, AC2, BEMS, NOr1, NOr2}. These include, in particular,  the well-known Harish-Chandra-Itzykson-Zuber (HCIZ) integral \cite{HC, IZ}, which plays a fundamental r\^ole both in representation theory and in coupled matrix models. In \cite{GGN1, GGN2, GH1}, it was shown that when the Toda flow parameters 
are equated to the trace invariants of a pair of $N \times N$  hermitian matrices, and the expansion parameter is equated to $-1/N$,  this gives the generating function for the enumeration of weakly monotonic paths in the Cayley graph with a fixed number of steps while, geometrically, it coincides with signed enumeration of  branched coverings of fixed genus and variable numbers of branch points \cite{HO3, GH2}.  Other matrix integrals
give ``hybrid'' paths consisting of both weakly and strongly monotonic segments or, equivalently, enumeration  of coverings with  multispecies  ``coloured'' branch points \cite{HO3, GH1}. Certain of these  may also be shown  to satisfy differential  constraints,  the so-called {\em Virasoro constraints} \cite{MM, Z, KZ}, due to reparametrization invariance, and {\em loop equations}  \cite{BM, BEMS, AC1, AC2}  following from the structure of the underlying matrix integrals.

These, and other generating functions for various enumerative,  topological, combinatorial and  geometrical invariants related to Riemann surfaces, such as intersection numbers \cite{Ko}, higher Gromov-Witten invariants,  Hodge numbers \cite{KL, K},  knot invariants \cite{BE, MMN},  and a growing number of other cases, can be placed into the  {\em topological recursion} scheme \cite{GJ2, EO1, EO2, EO3},  which  aims at  determining  the generating functions  through algorithmic recursion sequences stemming from  an underlying {\em spectral curve} \cite{BEMS, KZ}. This  has turned out to be a very effective approach to a broad class of  examples.

    However not all such generating functions are known to be $\tau$-functions in the usual sense of integrable systems, nor  partition functions or correlaters for matrix models. It remains something of a mystery  exactly which class of invariants is amenable to such a representation. A further remarkable fact  is that,  in some cases, different generating functions corresponding to  distinct enumerative problems, such as  Hurwitz numbers and  Hodge integrals, may  be $\tau$-functions that are related through algebraic transformations that themselves  involve the spectral curve \cite{K, KL}.
 
 The present work is concerned solely with the case of Hurwitz numbers, but in the generalized
 sense, allowing  infinite parametric families of weightings. It  provides a unified approach 
encompassing all  cases of weighted Hurwitz numbers that have appeared to date, 
interpreting these as special cases of an infinite parametric family of weighting functions
 determining mKP or 2D-Toda $\tau$-functions of  generalized hypergeometric type.
The parameters serve to specify the particular weighting used when summing over the various
configurations. Their values are determined by a ``weight generating'' function $G(z)$, and define the weighting by 
evaluation of the standard bases $(e_\lambda, h_\lambda, m_\lambda, f_\lambda)$, for the space 
$\Lambda$ of symmetric functions in an infinite number of indeterminates \cite{Mac} 
consisting of {\em elementary}, {\em complete}, {\em monomial} and {\em forgotten} symmetric functions, respectively, at the given set of parameters $(c_1, c_2, \dots)$ determined by $G(z)$.   The other two standard bases, the Schur functions  $\{s_\lambda\}$ and  the power sum symmetric functions $\{p_\mu\}$,  serve as bases for expansions of the $\tau$-function, in which  the coefficients in the first are diagonal and of {\em content product} form, guaranteeing that the Hirota bilinear equations of the integrable hierarchy are satisfied, while those in the second provide the weighted  Hurwitz numbers. 

 Besides the various ``classically weighted'' cases, arising through different choices of  the parameters
 $(c_1, c_2, \dots)$, there are also ``quantum deformations'', depending  on an additional pair  $(q,t)$ that
are closely linked to  the MacDonald symmetric functions \cite{Mac}.  This leads to the notion  of `quantum weighted''   Hurwitz numbers, of various types \cite{GH2, H2}, which may  depend  both on the  infinity of classical weighting parameters  $(c_1, c_2, \dots)$,  and the further  pair $(q,t)$,  in a specific way, involving $q$-deformations. 
 Another generalization consists of introducing multiple expansion parameters  $(z_1, z_2, \dots)$, leading to 
 generating functions for weighted ``multispecies'' weighted Hurwitz numbers \cite{H1},
 which are counted with different weighting factors, depending on the species type, or ``colour''. 

In \autoref{tau_functions}, a quick review is given of the fermionic approach to $\tau$-functions 
for the KP hierarchy and modified KP  sequence of $\tau$-functions as introduced by  Sato, 
\cite{Sa, SS} as well as the 2D  Toda case introduced in \cite{Ta, UTa, Takeb}. 
\autoref{center_group_algebra_hurwitz} recalls  basic notions regarding the $S_n$ group algebra, 
 including the commuting {\em Jucys-Murphy elements}\cite{Ju, Mu}, Frobenius' characteristic map from the center  
 $\Zb(\Cb[S_n])$   to the algebra $\Lambda$ of symmetric functions,  and the abelian group within  $\Zb(\Cb[S_n])$ that is generated through a combination of these. \autoref{hypergeom_hurwitz} gives a summary of the new approach to the construction of $\tau$-functions of  hypergeometric  type interpretable as generating functions for infinite parametric families of  weighted Hurwitz numbers  developed in \cite{GH1, GH2, H1, H2, HO3}. The weightings are interpreted both geometrically, as weighted enumeration of $n$-sheeted branched covers  of the Riemann sphere,  and combinatorially, as weighted enumeration of paths  in the Cayley graph of $S_n$ generated by transpositions.   The relation between these is  easily seen algebraically through the Cauchy- Littlewood generating functions for dual pairs of bases for $\Lambda$.

\autoref{examples_classical} is devoted to the various examples that have  so far been
considered in the literature. These include: the original case of single and double Hurwitz numbers, generated by the special KP and 2D Toda $\tau$-functions studied by Pandharipande and Okounkov \cite{Pa, Ok};  the case of the HCIZ integral \cite{IZ, HC, GGN1, GGN2, GH1}, which is known to have the combinatorial interpretation of counting  weakly monotonically increasing paths of transpositions in the Cayley graph, to which is added the geometrical one of signed enumeration of branched coverings with an arbitrary number of branch points with arbitrary branching profiles, at fixed genus; another case \cite{GH1}, which counts strongly monotonic such paths, and can be related to the special case of counting Belyi curves \cite{AC1, AC2, Z, KZ} (with three branch points) or ``Dessins d'enfants''; and
a  hybrid case \cite{GH1}, which combines the two, and counts branching configurations of multiple ``colour'' type and, moreover also has a matrix model representation. More general ``multispecies'' branched coverings, with their associated combinatorial equivalents \cite{HO3, H1}, and other, more general parametric families of weighted Hurwitz numbers are considered in \autoref{multispecies}.  

Already in the ``classical'' setting,   it is possible to select the parameters $(c_1, c_2, \dots)$ appearing in the associated weight generating functions in such a way that the resulting weightings, both for branched coverings and for paths, involve what may be interpreted as a quantum deformation parameter $q$ . When suitably interpreted in terms of Planck's constant  $\hbar$  and temperature, the resulting distributions can be related to the energy distribution  law for a Bose gas  with linear energy spectrum.  In \autoref{macdonald_quantum_hurwitz}, we extend the family of weight generating functions by  introducing  a further pair  $(q,t)$ of  deformation parameters that play the same r\^ole as those appearing in the MacDonald symmetric  functions \cite{Mac},  with the Cauchy-Littlewood generating functions replaced by the corresponding one for  Macdonald functions \cite{H2}.  The resulting weighted Hurwitz numbers are interpretable  as multispecies quantum Hurwitz numbers, whose distributions are again related to those for a Bosonic gas. Various specializations are obtained by choosing  specific values for the parameters $q$ and $t$, or relations between them, or various limits.  Besides  recovering the ``classical'' weighting, for $q=t$, 
this leads to various other specializations, such as weightings involving the quantum analog  of the elementary and complete symmetric functions,  the Hall-Littlewood polynomials and the Jack polynomials.


\subsection{Enumerative geometrical Hurwitz numbers}
For any set of partitions $\{\mu^{(1)}, \dots, \mu^{(k)}\}$  of $n \in \Nb^+$, we define the geometrical Hurwitz number 
$H(\mu^{(1)}, \dots, \mu^{(k)})$ to be the number of $n$-sheeted branched coverings of the Riemann
sphere having no more than $k$ branch points $\{q_1, \dots , q_k\}$, with ramification profiles 
of type $\{\mu^{(i)}\}$, weighted by the inverse of the order of their automorphism groups.
The Frobenius-Schur formula \cite{Frob1,  Frob2,  Sch,  LZ} expresses these in terms of the irreducible characters $\chi_\lambda(\mu^{(i)})$ of the symmetric  group $S_n$
\be
H(\mu^{(1)}, \dots, \mu^{(k)}) = \sum_{\lambda, |\lambda|=n} h_\lambda^{k-2} \prod_{i=1}^k  z_{\mu^{(i)}}^{-1} \chi_\lambda(\mu^{(i)})  
\label{frob_schur}
\ee
where $\lambda$ is the partition corresponding to the irreducible representation with Young symmetrizer of type $\lambda$,  and the parts of the partitions $\{\mu^{(i)}\}$ are the cycle lengths defining the ramifications profiles that determine the conjugacy classes $\cyc(\mu^{(i)})$ on which $\chi_\lambda$ is evaluated. Here
\be
z_\mu = \prod_{i=1}^{\ell(\mu^{(i)})} i^{m_i(\mu)} (m_i(\mu))!
\ee
is the order of the stabilizer  of any element in $\cyc(\mu)$ under conjugation,
where $m_i(\mu)$ is the number of parts of $\mu$ equal to $i$ and
 \be
h_\lambda := \det\left({1\over (\lambda_i - i+ j)!}\right)^{-1} 
\ee
is the product of the hook lengths of the partition $\lambda$.

\subsection{Combinatorial Hurwitz numbers}
The combinatorial definition of the Hurwitz number, denoted $F(\mu^{(1)}, \dots, \mu^{(k)})$  (and perhaps more
aptly called the {\em Frobenius number}, although the two turn out to be equal!) is the following: 
$n! F(\mu^{(1)}, \dots, \mu^{(k)})$ is the number of ways the identity  element $\Ib  \in S_n$ may be factorized into a product
\be
\Ib = g_1 \cdots g_k, 
\ee
in which  the $i$th factor $g_i\in S_n$ is in the conjugacy class $\cyc(\mu^{(i)})$.
  The equality of these two quantities 
   \be
  F(\mu^{(1)}, \dots, \mu^{(k)}) =H(\mu^{(1)}, \dots, \mu^{(k)})
  \label{hurwitz_frobenius}
  \ee
  follows from the monodromy representation of the fundamental group  $\pi_1(\Cb\Pb^1/\{q_1, \dots, q_k\})$ of the punctured sphere with the branch points removed  \cite[Appendix~A]{LZ}.
 
 As shown  in \autoref{hurwitz_numbers},  relation (\ref{frob_schur})  follows from  (\ref{hurwitz_frobenius})
and the Frobenius character formula. Avatars of this equality will be seen to recur repeatedly in the 
various versions of weighted  Hurwitz numbers studied below.

\section{mKP and 2D Toda $\tau$-functions}
\label{tau_functions}

\subsection{Fermionic Fock space}
\label{fermionic_fack_space}

The fermionic Fock space $\FF$ is defined \cite{SS} as the semi-infinite wedge product space
\be
\FF := \Lambda^{\infty/2} \HH 
\ee
constructed from a separable Hilbert space $\HH$ with orthonormal basis $\{e_i\}_{i\in \Zb}$,
that is split into an orthogonal direct sum of two subspaces
\be
\HH = \overline{\HH_+ \oplus \HH_-},
\ee
where
\be
\HH_- = \span \{e_i\}_{i\in \Nb}, \quad \HH_+=\span\{e_{-i}\}_{i\in \Nb^+}.
\ee
and $\{e_i\}_{i\in \Zb}$ is an orthonormal basis.

\br
The curious convention of using negative $i$'s to label the basis for $\HH_+$ and
positive ones for $\HH_-$ stems from the notion of the ``Dirac sea'',  in which
all negative energy levels are filled and all positive ones empty, where the
integer lattice is identified with the energies. If we take Segal and Wilson's \cite{SW} model
for $\HH$ 
\be
\HH := L^2(S^1) = \overline{\span\{z_i\}}_{i\in \Zb}  \text{ with }  e_i:= z^{-i-1},
\ee
we may view $\HH_+$ and $\HH_-$ either as the subspaces of  positive and negative Fourier series 
on the circle $S^1$ or, equivalently, the Hardy spaces of  square integrable functions admitting a 
holographic extension to inside and outside  the unit circle, with the latter vanishing at $z= \infty$.
\er

$\FF$ is the graded sum 
\be
\FF = \oplus_{N\in \Zb} \FF_N
\ee
of the subspaces $\FF_N$  with fermionic charge $N\in \Zb$. 
An orthonormal basis $\{|\lambda; N\rangle\}$ for these is provided by the
semi-infinite wedge product states
\be
|\lambda; N\rangle := e_{\ell_1} \wedge e_{\ell_2} \wedge \cdots
\ee
labeled by pairs of partitions $\lambda$ and integers $N\in \Zb$, where
\be
\{\ell_i := \lambda_i - i +N\}
\ee
are the ``particle coordinates'', indicating the occupied points on the integer lattice, corresponding to
the parts of the partition $\lambda$, with the usual convention that, for $i$ greater
than the length $\ell(\lambda)$ of the partition, $\lambda_i := 0$. The vacuum state in the
charge $N$ sector $\FF_N$ of the Fock space is denoted
\be
|N\rangle := | 0; N\rangle.
\ee

In Segal and Wilson's \cite{SW} sense,   the image $\PP(W)$,  under the Pl\"ucker map
\bea
\PP: Gr_{\HH_+}(\HH) &\&\ra \Pb(\FF) \cr
\PP: W\  &\&\ \mt \  \Pb(W) \cr
\PP: \Span\{w_i \in \HH\}_{i\in \Nb^+}  &\&\mapsto [w_1 \wedge w_2 \wedge \cdots],
\eea
of an element $W\in Gr_{\HH_+}(\HH)$ of the infinite  Grassmannian  modeled  
 on $\HH_+ \ss \HH$, having {\em virtual dimension}  $N$ (i.e., such that the  Fredholm
 index of the orthogonal projection map $\pi^{\perp}: W \ra \HH_+$ is $N$)
  is in the charge  $N$ sector $\PP(W)\in \FF_N \ss \FF$,  and the entire image  consists of all decomposable elements of $\FF$. 
 In particular,   $\HH_+$  is mapped  to the projectivization of the vacuum element
\be
\PP: \HH_+ \mt [|0 \rangle]:=[ |0;0\rangle] = [e_{-1}\wedge e_{-2}\wedge \cdots ] .
\ee

The Fermi creation and annihilation operators  $\psi_i$, $\psi_i^\dag$ are defined as exterior multiplication
by the basis element $e_i$ and interior multiplication by the dual basis element $\tilde e^i$, respectively.
\be
\psi :=  e_i \wedge  \quad \psi^\dag := i(\tilde{e}^i ).
\ee
These satisfy the usual anticommutation relations 
\be
[\psi_i, \,  \psi_j^\dag]_+ = \delta_{ij}
\ee
defining the corresponding Clifford algebra on $\HH + \HH^*$ with respect to the natural quadratic form
in which both $\HH$ and $\HH^*$ are totally isotropic.

The infinite general linear algebra $\grg\grl(\HH) \ss \Lambda^2(\HH +\HH^*)$, in the standard Clifford representation,
is spanned by the elements $:\psi_i \psi_j^\dag:$, with the usual convention for normal ordering
\be
:\psi_i \psi_j^\dag: = :\psi_i \psi_j^\dag:  - \langle \psi_i \psi_j^\dag \rangle,
\ee
where $\langle \OO\rangle$ denotes the vacuum expectation value
\be
\langle \OO\rangle := \langle 0 |\OO | 0\rangle.
\ee
The corresponding group  $\grG\grl(\HH)$  consists of invertible endomorphisms, 
having well defined determinants. (See \cite{SS, Sa, SW} for more detailed definitions.)

A typical exponentiated element in the Clifford representation is of the form
\be
\hat{g} = e^{\sum_{ij \in \Zb} A_{ij} :\psi_i \psi_j^\dag:},
\ee
where the doubly infinite square matrix with elements $A_{ij}$ satisfies suitable
convergence conditions \cite{Sa, SS, Sa, SW} that will not be detailed here.

\subsection{Abelian group actions,  mKP and 2D Toda lattice $\tau$-functions and Hirota relations}

     The KP and 2D-Toda flows are generated by the multiplicative action on $\HH$ of the two infinite abelian subgroups
     $\Gamma_{\pm} \ss \grG\grl_0(\HH)$ of the identity component $\grG\grl_0(\HH)$ of the general linear group $\grG\grl(\HH)$,
     defined by:
 \be
  \Gamma_+:=\{\gamma_+({\bf t}):=e^{\sum_{i=1}^\infty t_i z^i}\},
 \quad {\rm and \quad}  \Gamma_- :=\{\gamma_-({\bf s}):=e^{\sum_{i=1}^\infty s_i z^{-i}}\},
 \ee
where  ${\bf t}= (t_1, t_2, \dots)$  is an infinite sequence of (complex) flow parameters corresponding to one-parameter 
subgroups, and  ${\bf s}= (s_1, s_2, \dots)$ is a second such sequence.   These in turn have the following Clifford group
representations on $\FF$
\be
\hat{ \Gamma}_+:=\{\hat{\gamma}_+({\bf t}):=e^{\sum_{i=1}^\infty t_i J_i}\},
 \quad {\rm and \quad}  \hat{\Gamma}_- :=\{\hat{\gamma}_-({\bf s}):=e^{\sum_{i=1}^\infty s_i J_{-i}}\},
\ee
where
\be
J_i:= \sum_{k \in \Zb}: \psi_i \psi_{i+k}^\dag :, \quad \pm i \in \Nb^+
\ee
are referred to as the ``current components''. In this infinite dimensional setting, whereas the abelian
groups $\Gamma_\pm$ commute, their Clifford representations $\hat{\Gamma}_\pm$ involve
a central extension, so that
\be
\hat{\gamma}_+({\bf t})\hat{\gamma}_-({\bf s}) = \hat{\gamma}_-({\bf s}) \hat{\gamma}_+({\bf t}) e^{\sum_{i\in \Zb} i t_i s_i}.
\ee
    
      The mKP-chain  and 2D-Toda $\tau$-functions corresponding to the element
$g  \in  \grG\grl_0(\HH$ are given, within a nonzero multiplicative constant, by the vacuum expectations values (VEV's)
       \bea
    \tau_g^{mKP} (N, {\bf t})&\& := \langle  N \vert  \hat{\gamma}_+({\bf t})  \hat{g} \vert N \rangle,
    \label{KPtau} \\
    \tau^{(2Toda)}_g(N, {\bf t}, {\bf s}) &\&: =    
    \langle N \vert  \hat{\gamma}_+({\bf t})\hat{g}\hat{\gamma}_-({\bf s}) \vert N\rangle.
 \label{2KPtau}
  \eea
    If the group element $g\in  \grG\grl_0(\HH $ is interpreted, relative to the  $\{e_i\}_{i\in \Zbb}$ basis, as a 
    matrix exponential  $g=e^A$, where the algebra element $A\in \grgl(\HH)$ is represented by the  infinite matrix with 
    elements $\{A_{ij}\}_{i,j \in \Zbb}$, then  the corresponding representation of $GL(\HH)$ on $\FF$ is given by
   \be
    \hat{g} := e^{\sum_{i,j \in \Zbb} A_{ij} : \psi_i \psi^\dag_j : } ,
      \label{ghat}
\ee

These satisfy the Hirota bilinear relations
\bea
&\& \oint_{z=\infty} z^{N'-N}e^{-\xi(\delta{\bf t}, z)} \tau^{mKP}_g(N, { \bf t}+ \delta{\bf t}  + [z^{-1}]) 
\tau^{mKP}_g(N',{ \bf t} -[z^{-1}] ) =0
 \label{hirotaKP} 
\\
 &\&  \oint_{z=\infty}  z^{N'-N}e^{-\xi(\delta{\bf t}, z)} \tau^{2D\text{T}}(N,{ \bf t} +[z^{-1}] , {\bf s})
 \tau^{2D\text{T}}(N', { \bf t}+ \delta{\bf t}  - [z^{-1}],  {\bf s} + \delta{\bf s})  = 
 \cr
 &\&  \oint_{z=0}  z^{N'-N} e^{-\xi(\delta{\bf s}, z^{-1})} \tau^{2D\text{T}}(N-1, {\bf t},   { \bf s} +[z])
 \tau^{2D\text{T}}(N'+1, {\bf t} + \delta{\bf t}, {\bf s}+ \delta{\bf s}  - [z] ) 
\label{hirotaToda}
 \eea
understood to hold  identically in  $\delta{\bf t}   = (\delta t_1, \delta t_2, \dots),\   \delta{\bf s} := (\delta s_1, \delta s_2, \dots)$,
where
\be
 \quad  [z]_i  :=  {1 \over  i }z^i.
  \ee

\subsection{Bose-Fermi equivalence and Schur function expansions}
\label{bose_fermi_schur_expansion}

   It follows from the identities \cite{Sa, SS}
   \be
   \langle N | \hat{\gamma}_+({\bf t}) | \lambda; N\rangle =  \langle\lambda; N | \hat{\gamma}_-({\bf t}) | N\rangle = s_\lambda({\bf t}), 
   \ee
   where $s_\lambda$ is the Schur function corresponding to partition $\lambda$, viewed
   as  function of the parameters
   \be 
   t_i := {p_i \over i},
   \ee
   where the $p_i$'s are the power sums, that the $\tau$-functions may be expressed,
   at least formally, as single and double Schur functions expansions
  \bea
  \tau_g^{mKP}(N, {\bf t}) &\& = \sum_{\lambda} \pi_\lambda(N, g) s_\lambda ({\bf t})  \\
   \tau_g^{2Toda}(N, {\bf t}, {\bf s})) &\& = \sum_{\lambda} \sum_\mu B_{\lambda \mu} (N, g) s_\lambda ({\bf t}) s_\mu ({\bf t})
  \eea
  where
  \be
  \pi_\lambda(N,g) :=  \langle \lambda; N | \hat{g} |N \rangle, \quad   B_{\lambda \mu}(N, g) := \langle \lambda; N | \hat{g} |\mu; N \rangle,
    \ee
  are the Pl\"ucker coordinates of the elements $ \hat{g} |N \rangle$ and $\hat{g} |\mu; N\rangle$ when $g\in \grG\grl_0(\HH)$ is in the
  identity component of $\grG\grl(\HH)$. The Hirota bilinear relations (\ref{hirotaKP}), (\ref{hirotaToda}) are then equivalent to the Pl\"ucker  relations satisfied by these coefficients. 
  
  The ``Bose-Fermi equivalence'' gives an isomorphism between  a completion $\BB_0$ of the space of   symmetric functions $\Lambda$  of an infinite number of ``bosonic'' variables $\{x_i \}_{i\in \Nb}$, labelled by the natural numbers  and the $N=0$ (zero charge) sector of the Fermionic Fock space $\FF_0 \ss \FF$ which identifies the basis  states $\{|\lambda;  0 \rangle\}$ with the basis of Schur functions $\{s_\lambda \in \Lambda\}$  through the ``bosonization'' map:
  \bea
 \Bb : \FF_0 \, &\& \ra \, \BB_0 \cr
 \Bb: | v \rangle  \ &\& \  \mt  \  \langle 0 | \hat{\gamma}_+ |v\rangle \cr
 \Bb: |\lambda; 0 \rangle  \ &\& \ \mt \ \ s_\lambda.
  \eea
  More generally, this can be extended to the full (graded) fermonic Fock space $\FF= \oplus_{N \in \Zb}\FF_N$
  by adding a  parameter $\zeta$ to the Bosonic Fock space, taking  formal Laurent expansions in this 
\be
\BB:= \BB_0 [[\zeta]],
\ee
and defining
\bea
 \Bb : \FF_N  &\&  \ra \, \BB_N \cr
 \Bb: | v \rangle  &\& \ \mt \  \langle N | \hat{\gamma}_+ |v\rangle  \zeta^N
 \eea
 
 Using $\Bb$ as an intertwining map, this defines identifications between
 operators in $\End(\FF)$ and those in $\End(\BB)$. However, what appears
 in the Fermonic representation as a ``locally'' defined element  of the Clifford algebra
 or group is in general  a nonlocal operator in the Bosonic representation (involving exponentiated  
 differential operators  in terms of the ${\bf t}$ coordinates), as is the case, e.g. , for the
 Bosonic representations of the operators $\psi, \psi_i^\dag $, which are special types
 of ``vertex operators''. In particular, the Bosonization of fermionic states of the type
 $\hat{g} \hat{\gamma}_- | 0\rangle$ is given by  application of  nonlocal operators of the type
  that were interpreted in \cite{GJ1}   as ``cut-and-join'' operators, to the gauge transform 
  of the vacuum state, defined by $ \hat{\gamma}_- | 0\rangle$
 
\subsection{Hypergeometric $\tau$-functions and convolution symmetries}
\label{hypergeometric_convolution_sym}
\label{convolution_symmetries}

A special subfamily of the above consists of those $\tau$-functions for which the group
element $\hat{g}$ is diagonal 
\be
\hat{g} = e^{\sum_{i \in \Zb} T_i : \psi_i \psi_i^\dag}, \quad A_{ij} = T_i \delta_{ij}
\label{conv_symm}
\ee
in the basis $|\lambda;N\rangle$. These were named {\em convolutions symmetries} in \cite{HO2}, since in the Segal-Wilson representations of $\grG\grl(\HH)$ they may be interpreted as (generalized) convolution products
on $\HH \sim L^2(S_1)$. Their eigenvalues $ r_\lambda(N, g)$ in the basis $|\lambda;N\rangle$
\be
e^{\sum_{ i \in \Zb} T_i : \psi_i \psi_i^\dag } |\lambda; N\rangle = r_\lambda(N, g) |\lambda; N\rangle
\ee
 can be written in the form of a {\em content product} \cite{OrSc, HO2}:
\be 
r_\lambda(N, g) := r_0(N,g)  \prod_{(i,j) \in \lambda} r_{N+j-i}( g), \quad r_i(g) := e^{T_{i} - T_{i-1}}
\label{content_product_fermion}
\ee
where
\be
r_0(N,g) := \begin{cases} \prod_{i=0}^{N-1}e^{T_i} \quad {\rm if} \quad N > 0 \cr
           \quad  \ 1 \qquad \ \     {\rm if} \quad  N=0  \cr
        \prod _{i=N}^{-1} e^{-T_i} \quad \ \  {\rm if } \quad N < 0.
        \end{cases}
\ee
The double Schur function expansion (\ref{2KPtau}) in this case reduces to the diagonal form
 \be
   \tau_g^{2Toda}(N, {\bf t}, {\bf s}))  = \sum_{\lambda}  r_\lambda(N,g) s_\lambda ({\bf t}) s_\lambda ({\bf s}).
     \label{2DT_hypergeom} 
  \ee
  
  If we view the second set of parameters $(c_1, c_2, \dots)$ as fixed, and consider only the first set $(1_1, 1_2,, \dots)$ as KP flow parameters, we may interpret (\ref{2DT_hypergeom}) as defining a chain of mKP $\tau$ functions.
  A specific value of special interest is $(c_1, c_2, \dots )$ = $(1, 0, 0 \dots)$, for which the Schur function
  evaluates to
  \be
  s_\lambda(1,0, \dots) = h_\lambda^{-1}
  \ee
  and  (\ref{2DT_hypergeom}) reduces to 
  \be
   \tau_g^{mKP}(N, {\bf t}, {\bf s}))  = \sum_{\lambda}  r_\lambda(N,g) h_\lambda^{-1}s_\lambda ({\bf t}).
     \label{mKP_hypergeom}
  \ee

  In the following, only such {\em hypergeometric} $\tau$-functions will be needed.
 By defining suitable parametric families of the latter, and expanding these in powers
of some auxiliary parameters, while leaving the others to define the weightings,  It will be seen that
we can interpret  them as generating functions  for finite or  infinite parametric families of weighted Hurwitz numbers, both classical and quantum, obtaining both a natural enumerative geometric and combinatorial interpretation in all cases.

\section{The center $\Zb(\Cb[S_n])$ of the $S_n$ group algebra and symmetric functions}
\label{center_group_algebra_hurwitz}

\subsection{The $\{C_\mu\}$ and $\{F_\lambda\}$ bases}

There are two natural bases for the center $\Zb(\Cb[S_n])$ of the group algebra of
the symmetric group $S_n$, both labelled by partitions of $n$. The first  is
the basis of cycle sums $\{C_\mu\}|_{|\mu|=n}$, defined by
\be
C_\mu := \sum_{h\in \cyc(\mu)} h.
\ee
The second is the basis of orthogonal idempotents $\{F_\lambda\}_{|\lambda|=n}$,
which project onto the irreducible representations of type $\lambda$ and satisfy
\be
F_\lambda F_\mu = F_\lambda \delta_{\lambda \mu}.
\label{FF_F}
\ee
These are related by
\bea
F_\lambda &\&=  h_\lambda^{-1}\sum_{\mu, \, |\mu|=|\lambda| =n} \chi_\lambda(\mu) C_\mu
\label{F_lambda_C_mu}
\\
C_\mu &\&= z_\mu^{-1}\sum_{\lambda, \, |\lambda| = |\mu| =n} h_\lambda \chi_\lambda(\mu) F_\lambda
\label{C_mu_F_lambda}
 \eea
 which is equivalent to the Frobenius character formula (see below).
 The main property of the $\{F_\lambda\}$ basis is that multiplication by any element
 of the center $\Zb(\Cb[S_n])$ is diagonal in this basis (as follows immediately from
 (\ref{FF_F})).

\subsection{The characteristic map}

Frobenius'  characteristic map defines a linear isomorphism between the characters
of $S_n$ and the characters of tensor representations of $GL(k)$, of total tensor
weight $n$, for $k$ sufficiently large. It maps the irreducible character $\chi_\lambda$ to the
Schur function $s_\lambda$, viewed as the corresponding $GL(k)$ character through the
Weyl character formula for any  $k\ge \ell(\lambda).$ Equivalently, it defines a linear endomorphism 
\bea
\ch: \Zb(\Cb[S_n]) &\& \, \ra \, \Lambda \cr
\ch:  F_\lambda  &\& \ \,  \mt \ {s_\lambda \over h_\lambda}
\eea
from the  center $\Zb(\Cb[S_n])$  of the group algebra to the algebra $\Lambda$ of symmetric functions \cite{Mac}.
The change of basis formulae (\ref{F_lambda_C_mu}), (\ref{C_mu_F_lambda}), together
with the Frobenius character formula
\be
s_\lambda = \sum_{\mu, \, |\mu|= |\lambda| =n} z_\mu^{-1} \chi_\lambda(\mu) p_\mu, 
\ee
where
\be
p_\mu := \prod_{i=1}^{\ell(\mu)} p_{\mu_i}
\ee
is the power sum symmetric function, then imply  that the characteristic map
takes the cycle sum basis into the $\{p_\mu\}$ basis for $\Lambda$
\be
\ch: C_\mu \, \mt \,  {p_\mu \over z_\mu}. 
\ee

\subsection{Combinatorics of Hurwitz numbers and the Frobenius-Schur formula}
\label{hurwitz_numbers}

The two bases $\{C_\mu\}$,  $\{F_\lambda\}$ can be used to deduce
the Frobenius-Schur formula (\ref{frob_schur}),  expressing $H(\mu^{(1)}, \dots, \mu^{(k)})$
in terms of the irreducible group characters $\chi_\lambda(\mu)$.
The  product $\prod_{i=1}^k C_{\mu^{(i)}}$ of elements of the cycle sum basis
is central  and hence can be expressed relative to the same basis:
\be
\prod_{i=1}^k C_{\mu^{(i)}} = \sum_{\nu, |nu|= n} H(\mu^{(1)}, \dots, \mu^{(i)}, \nu) z_\nu C_\nu,
\label{C_mu_prod}
\ee
and, in particular, the coefficient of the identity class, for which $\mu = (1)^n$ is $n!$ times
the Hurwitz number
\be
[\Ib = C_{ (1)^n}]\prod_{i=1}^k C_{\mu^{(i)}}= n! H(\mu^{(1)},  \dots , \mu^{(k)}), 
\label{C_mu_prod_Id}
\ee
giving the number of factorizations of the identity element into a product of
$k$ elements within the conjugacy classes $\{\cyc(\mu^{(i)}\}_{i=1 , \dots, k}$.

Substituting  the change of basis formula (\ref{C_mu_F_lambda}) into (\ref{C_mu_prod_Id}),
applying both sides to the basis element $\{F_\lambda\}$ and equating the eigenvalues
that result gives the Frobenius-Schur formula:
\be
H(\mu^{(1)},  \dots , \mu^{(k)}) = \sum_{\lambda, \, |\lambda|=|\mu|= n} h_\lambda^{k-2} \prod_{i=1}^k {\chi_\lambda(\mu^{(i)} )\over z_\mu^{(i)}}. 
\ee

\subsection{Jucys-Murphy elements, central elements and weight generating functions }

We now recall the special commuting elements $(\JJ_1, \dots, \JJ_n\}$ of the group algebra $\Cb[S_n]$ introduced
by Jucys \cite{Ju} and  Murphy \cite{Mu}. (See also \cite{DG}). These are defined by
\be
\JJ_b := \sum_{a=1}^{n-1} (ab) \  \text{ for } b>1, \text{ and }  J_1 :=0.
\ee
where $(ab) \in S_n$ is the transposition that interchanges $a$ with $b$.
Although these are not central elements, they have two remarkable properties:
Any symmetric function $f(\JJ_1, \dots , \JJ_n)$, $f \in \Lambda_n$ formed from them is
central, and this central element has eigenvalues in the $F_\lambda$ basis that
are equal to the evaluation on the {\em content} of the partition $\lambda$; i.e. the
set of number $j-i$, where $\{(i,j) \in \lambda\}$  are the set of positions (in the English
convention) in the Young diagram  of $\lambda$:
\be
f(\JJ_1, \dots, \JJ_n) F_\lambda = f(\{j-i\}_{(ij) \in \lambda}) F_\lambda.
\ee
  
  A particular case of symmetric functions of $n$ variables consists of taking
 a single generating function $G(z)$,  expressed formally either as an infinite product
 \be
 G(z) = \prod_{i=1}^\infty (1+ c_i z)
 \label{G_inf_prod}
 \ee
 or an infinite sum
 \be
 G(z) = 1 +\sum_{i=1}^\infty G_i z^i
 \ee
 or some limit thereof, and defining the central element as a product
 \be
 G_n(z, \JJ) :=\prod_{b=1}^n G(z\JJ_a).
 \ee
    (For the present, we are not concerned with whether $G(z)$
 is polynomial, rational,  a convergent series,  in some field extension or just a formal 
 infinite series or infinite product;   the considerations that  follow are mainly algebraic,  but are easily extended 
  to include either convergent series, through suitable completions,  or formal series and products,
  as in the generating functions for symmetric functions.)
  When applied multiplicatively to the $\{F_\lambda\}$  basis,  the  central element $G_n(z, \JJ)$  has eigenvalues
  that are expressible as content products
  \be
   G_n (z, \JJ ) F_\lambda= \prod_{(ij) \in \lambda} G(z(j-i)) F_\lambda, \quad |\lambda| =n .
   \label{content_product_center}
  \ee

    We also consider the ``dual'' generating function:
 \be
 \tilde{G}(z) := {1\over G(-z)} = \prod_{i=1}^\infty (1 - c_i z)^{-1} 
  \label{tilde_G_inf_prod}
 \ee
 and associated central element
 \be
 \tilde{G}_n(z, \JJ ) :=\prod_{b=1}^n \tilde{G}(z\JJ_a),
 \ee
 which similarly satisfies
  \be
   \tilde{G}_n (z, \JJ ) F_\lambda= \prod_{(ij) \in \lambda} \tilde{G}(z(j-i)) F_\lambda, \quad |\lambda| =n .
   \label{dual_content_product_center}
  \ee
 
  This  suggests comparison with the ``convolution symmetry'' elements in the fermionic representation
  of the group $\grG \grl(\HH)$ and an extension of the Bose-Fermi equivalence, using the
  characteristic map, to a correspondence between the direct sum $\oplus_{n\in \Nb} \Zb(\Cb[S_n])$
  and the $N=0$ sector $\FF_0\ss \FF$ of the fermonic Fock space.
  
\subsection{Bose-Fermi equivalence and  $\oplus_{n\in \Nb} \,\Zb(\Cb[S_n])$}
\label{hurwitz_numbers}

Composing the characteristic map with the Bose-Fermi equivalence we obtain
an endomorphism $\EE$ from the direct sum $\oplus_{n \in \Nb} \,\Zb(\Cb[S_n])$ of the  centers of
the group algebras to the zero charge sector $\FF_0$ in the Fermionic Fock space
\bea
\EE : \oplus_{n\in \Nb}\, \Zb(\Cb[S_n])\,  &\& \ra \,  \FF_0 \cr
\EE: F_\lambda \  &\& \   \mt  \ h_\lambda^{-1} | \lambda; 0\rangle
\label{EE_def}
\eea
This provides an intertwining map between the central elements in the completion of
the group algebra formed from products of functions of a single variable, acting
by multiplications, and the convolution symmetries discussed in \autoref{hypergeometric_convolution_sym}.

Choosing the parameters $T_j$ in (\ref{conv_symm}) as
\be
T^{G(z)}_j = \sum_{k=1}^j \ln G(zk), \quad T^{G(z)}_0(z) = 0, \quad T^{G(z)}_{-j}(z) = -\sum_{k=0}^{j-1} \ln G(-zk) \quad\text{for $j>0$}.
\ee
so that
\be
\hat{g} =\hat{C}_G := e^{\sum_{i \in \Zb} T^G_i (z): \psi_i \psi_i^\dag:} ,
\label{conv_symm_G}
\ee
it follows that
\be
r_j(g) := r_j^{G(z)}= G(jz)
\ee
and
\be
\hat{C}_G \ket{\lambda; N} = r_\lambda^{G(z)}(N) \ket{\lambda; N}
\label{CG_lambda}
\ee
with eigenvalues
\be
r_\lambda^{G(z)}(N) \deq r^{G(z)}_0(N) \prod_{(i,j)\in \lambda} G(z(N+ j-i)),
\ee
where
\be
r_0^{G(z)}(N) = \prod_{j=1}^{N-1} G((N-j)z)^j, \quad r_0(0)  = 1, \quad r_0^{G(z)}(-N) = \prod_{j=1}^{N} G((j-N)z)^{-j},
\quad N>1
\ee

The map $\EE$ defined in ({\ref{EE_def}) therefore intertwines the action
of $\oplus_{n\in \Nb}G_n(z, \JJ)$ on $\oplus_{n\in \Nb} \,\Zb(\Cb[S_n])$ with that of $\hat{C}_G $ on $\FF_0$.
The same applies to the dual generating functions $\tilde{G}(z)$, for which we obtain
the corresponding content product formula expression
\be
r_\lambda^{\tilde{G}(z)}(N) \deq r^{\tilde{G}(z)}_0(N) \prod_{(i,j)\in \lambda} \tilde{G}(z(N+ j-i)).
\ee
For the following, we only have need of the $N=0$ case, for which we simplify the notation 
for the content product coefficients to
\bea
r_\lambda^{G(z)} &\& \deq r_\lambda^{G(z)}(0) =  \prod_{(i,j)\in \lambda} G(z( j-i)), \\
r_\lambda^{\tilde{G}(z)} &\& \deq r_\lambda^{\tilde{G}(z)}(0) =  \prod_{(i,j)\in \lambda} \tilde{G}(z( j-i)), 
\eea

\section{Hypergeometric $\tau$-functions as generating functions for weighted Hurwitz numbers }
\label{hypergeom_hurwitz}

We are now ready to state the main results, which show that the
KP  and 2D Toda $\tau$-functions of hypergeometric type
\bea
  \tau^{G(z)}({\bf t}) &\& = \sum_{\lambda} r^{G(z)}_\lambda h_\lambda^{-1}s_\lambda ({\bf t}),   \\
   \tau^{G(z)} ({\bf t}, {\bf s})) &\& = \sum_{\lambda} \ r^{G(z)}_\lambda s_\lambda ({\bf t}) s_\lambda ({\bf s}),
  \eea
 when expanded in bases of (products of) the power sum symmetric functions $\{p_\mu\}$, are  
 interpretable as generating functions for suitably defined infinite parametric weighted Hurwitz numbers, both in the enumerative geometric and the combinatorial sense.   The details and proofs  may be found in \cite{GH1, GH2, H1, H2, HO3}.

\subsection{The Cauchy-Littlewood formula and dual bases for $\Lambda$}

We have already encountered the two bases consisting of Schur functions $\{s_\lambda\}$ and power sum symmetric
functions $\{p_\lambda\}$ for the ring $\Lambda$ of symmetric functions in an arbitrary number
of indeterminates \cite{Mac}. In addition to these, there are four other useful bases, consisting of the products of the elementary symmetric functions 
\be
e_\lambda ({\bf x}) := \prod_{i=1}^{\ell(\lambda} e_{\lambda_i}
\ee
the complete symmetric functions 
\be
h_\lambda( {\bf x} ):= \prod_{i=1}^{\ell(\lambda} h_{\lambda_i},
\ee
with generating functions
\be 
E(z) = \prod_{ij} (1+z x_i) = \sum_{i=0}^\infty e_i z^i, \quad H(z) = \prod_{ij} (1-z x_i)^{-1} = \sum_{i=0}^\infty h_i z^i,
\ee
the monomial sum symmetric functions
\be
m_\lambda ({\bf x}) :=
\frac{1}{\abs{\aut(\lambda)}} \sum_{\sigma\in S_k} \sum_{1 \le i_1 < \cdots < i_k}
 x_{i_\sigma(1)}^{\lambda_1} \cdots x_{i_\sigma(k)}^{\lambda_k},
\ee
and the ``forgotten'' symmetric functions
\be
f_\lambda ({\bf x}) :=
\frac{(-1)^{\ell^*(\lambda)}}{\abs{\aut(\lambda)}} \sum_{\sigma\in S_k} \sum_{1 \le i_1 \le \cdots \le i_k} 
x_{i_\sigma(1)}^{\lambda_1},  \cdots x_{i_\sigma(k)}^{\lambda_k},
\ee
where
\be
{\bf x} := (x_1, x_2, \dots )
\ee
is an infinite sequence of indeterminates,  and defining $m_i(\lambda)$  to be the number of parts of $\lambda$ 
equal to $i$,
\be
\abs{\aut(\lambda)} := \prod_{i=1}^{\ell(\lambda)}(m(\lambda_i))! 
\ee
 is the order of the automorphism group of the conjugacy class of type $\lambda$ under conjugation.

These bases have the following duality and orthogonality properties with respect to the standard 
scalar product $(\, , \, )$ in which the Schur functions are orthonormal \cite{Mac}:
\be
(s_\lambda, s_\mu) = \delta_{\mu \nu}, \quad (p_\lambda, p_\mu) = z_\mu  \delta_{\mu \nu},
\quad (e_\lambda, m_\mu) = \delta_{\mu \nu}, \quad \quad (f_\lambda, h_\mu) = \delta_{\mu \nu}.
\ee
It follows \cite{Mac} that the Cauchy-Littlewood formula is expressible bilinearly in terns of l these dually paired bases
\bea
\prod_{i=1}^\infty\prod_{j=1}^\infty (1-x_i y_j) ^{-1}&\&= \sum_\lambda s_\lambda({\bf x}) s_\lambda({\bf y}) \\
&\&= \sum_\lambda z_\mu^{-1}p_\lambda({\bf x}) p_\lambda({\bf y}) \\
&\&= \sum_\lambda e_\lambda({\bf x}) m_\lambda({\bf y})=   \sum_\lambda e_\lambda({\bf y}) m_\lambda({\bf x})\\
&\&= \sum_\lambda f_\lambda({\bf x}) h_\lambda({\bf y}) = \sum_\lambda f_\lambda({\bf y}) h_\lambda({\bf x}) .
\eea
The dual Cauchy-Littlewood generating function is similarly expressed in terms these in a dual way \cite{Mac}:
\bea
\prod_{i=1}^\infty\prod_{j=1}^\infty (1+x_i y_j) &\&= \sum_\lambda s_\lambda({\bf x}) s_{\lambda'}({\bf y}) \\
&\&= \sum_\lambda (-1)^{\ell^*(\lambda)}z_\mu^{-1}p_\lambda({\bf x}) p_\lambda({\bf y}) \\
&\&= \sum_\lambda e_\lambda({\bf x}) f_\lambda({\bf y})=   \sum_\lambda e_\lambda({\bf y}) f_\lambda({\bf x})\\
&\&= \sum_\lambda h_\lambda({\bf x}) m_\lambda({\bf y}) = \sum_\lambda h_\lambda({\bf y}) m_\lambda({\bf x}) ,
\eea
where $\lambda'$ is the partition whose Young diagram is the transpose  of that for $\lambda$ and
\be
\ell^*(\lambda) = |\lambda| -\ell(\lambda)
\ee
is the {\em colength} of $\lambda$ (i.e., the complement of its length).

We make use of these formulae in the following way: for the indeterminates $(x_1, x_2, \dots )$
we substitute the parameters $(c_1, c_2, \dots )$ defining the weight generating function $G(z)$ 
as an infinite product, or its dual $\tilde{G}(z)$, while for the indeterminates $(y_1, y_2, \dots )$,
we substitute the $x$ times the Jucys-Murphy elements $(z\JJ_1, z\JJ_2,  \dots )$ up to a finite number $n$ of
these, and $0$ for the rest, to obtain a finite sum in $j$.

This gives the following central elements, expressed as sums of products of these bases,
either evaluated on the contents $(c_1, c_2, \dots )$ or the commuting elements $(\JJ_1, \JJ_2, \dots)$:
\bea
G_n(z,\JJ) &\&= \prod_{i=1}^n \prod_{a=1}^n(1 + zc_i \JJ_a)\\
&\&= \sum_{d=0}^\infty z^d\sum_{\lambda, \, |\lambda|=d} e_\lambda({\bf c}) m_\lambda(\JJ) 
\label{G_e_lambda_m_lambda}
\\
&\&=  \sum_{d=0}^\infty z^d\sum_{\lambda, \, |\lambda|=d}m_\lambda({\bf c}) e_\lambda(\JJ) 
\label{G_m_lambda_e_lambda}
\eea
and 
\bea
\tilde{G}_n(z,\JJ) &\&= \prod_{i=1}^n \prod_{a=1}^n(1 - zc_i \JJ_a)^{-1} \\
&\& =\sum_{d=0}^\infty z^d \sum_{\lambda, \, |\lambda|=d} h_\lambda({\bf c}) m_\lambda(\JJ)
\label{tilde_G_h_lambda_m_lambda}
 \\
&\&= \sum_{d=0}^\infty z^d \sum_{\lambda, \, |\lambda|=d}  f_\lambda({\bf c}) e_\lambda(\JJ) .
\label{tilde_G_f_lambda_e_lambda}
\eea

Recall that the elements $G_n(z,\JJ) , \tilde{G}_n(z,\JJ) \in \Zb(\Cb[S_n])$ are diagonal
in the basis of orthogonal idempotents, with the contact product coefficients as eigenvalues
\be
G_n(z,\JJ) F_\lambda = r_\lambda^{G(z)} F_\lambda, \quad 
\tilde{G}_n(z,\JJ) F_\lambda = r_\lambda^{\tilde{G}}(z) F_\lambda
\ee
where
\bea
r_\lambda^{G(z)} &\&= \prod_{(ij) \in \lambda} \prod_{k=1}^\infty (1 + z c_k (j-i)) \\
r_\lambda^{\tilde{G}(z)} &\&= \prod_{(ij) \in \lambda} \prod_{k=1}^\infty (1 - z c_k (j-i))^{-1}.
\eea

\subsection{Multiplication by $m_\lambda(\JJ) $  and  $e_\lambda(\JJ) $ in the $C_\mu$ basis}

In order to proceed further, we need to compute the effect of multiplication 
by $G_n(z,\JJ)$ and $\tilde{G}_n(z,\JJ) $ in the basis $\{C_\mu\}$ of $\Zb(\Cb[S_n])$ consisting
of cycle sums. Combinatorially, this requires the notion of the {\em signature} of a path in
the Cayley graph of $S_n$ generated by the transpositions.
 \begin{definition} 
 Given a $d$-step path in  the Cayley graph of $S_n$ generated by transpositions $(ab)$, 
 $a, b \in \{1,   \dots, n \}$, $a<b$,  consisting of the sequence:
 \be
 h, \ra (a_1 b_1)h \ra   (a_2 b_2) (a_1 b_1) h\ra \dots  \ra (a_d b_d) \cdots (a_1 b_1) h,
 \ee
 its {\em signature} $\lambda$ is the partition of weight $|\lambda|=d$  whose length $\ell(\lambda)$ equals the number of distinct second elements  appearing in the sequence, and whose parts $\{ \lambda_i\}_{i =1, \dots, d}$ consist of the number of  times each given second element occurs. 
 \end{definition}

  The  effect of multiplication of  $C_\mu$ by  the central element  $m_\lambda(\JJ) \in \Zb(\Cb[S_n])$
  is given by the following easily proved lemma  \cite{GH2}:
  \begin{lemma}
  \label{m_lambda_JJ_C_mu}
\label{generating_weighted_paths}
Multiplication by $m_\lambda(\JJ)$ defines an endomorphism of $\ZZ(\Cb[S_n])$ which, expressed in the $\{C_\mu\}$ basis, is given by
\be
m_\lambda(\JJ) C_\mu = {1\over |\mu|!} \sum_{\nu, \, \abs{\nu}=\abs{\mu}} m^\lambda_{\mu \nu} z_\nu C_\nu,
\ee
where $m^\lambda_{\mu \nu}$ is the number of monotonic $\abs{\lambda}$-step paths in the Cayley graph of $S_n$ generated by all transpositions, starting from  an element $h$ in the conjugacy class $\cyc(\nu)$ to $\cyc(\mu)$ with signature $\lambda$. Equivalently, 
\be
m^\lambda_{\mu \nu} :=
 \frac{\prod_{i=1}^{\ell(\lambda)} \lambda_i!}{\abs{\lambda}!} \, \tilde{m}^\lambda_{\mu \nu}
 \label{m_lambda_C_mu}
\ee
where $\tilde{m}^\lambda_{\mu \nu} $ is the number of  $|\lambda|$ step paths of signature $\lambda$ in the Cayley graph of $S_n$ generated by transpositions,  starting at the conjugacy class $\cyc{\mu}$ and ending in  the class $\cyc(\nu)$  

\end{lemma}

On the other hand, the  effect of multiplication of  $C_\mu$ by  the central element 
  $e_\lambda(\JJ) \in \Zb(\Cb[S_n])$  is given by the following \cite{GH2}:

\begin{lemma}
Multiplication by $e_\lambda(\JJ)$ defines an endomorphism of $\ZZ(\Cb[S_n])$ which, expressed in the $\{C_\mu\}$ basis, is given by
 \label{e_lambda_JJ_C_mu}
\label{generating_weighted_covers}
\bea
e_\lambda(\JJ) C_\mu &\& = \sum_{\substack {\mu^{(1)}, \dots , \, \mu^{(k)} , \\ \{\ell^*(\mu^{(i}) = \lambda_i\}} }
\left(\prod_{i=1}^k C_{\mu^{(i)}} \right)C_\mu
\label{e_lambda_C_mu} \\
&\& = \sum_{\substack {\mu^{(1)}, \dots , \, \mu^{(k)} , \\ \{\ell^*(\mu^{(i}) = \lambda_i\}} }
H(\mu^{(1)}, \dots, \mu^{(k)}, \mu, \nu) z_\nu C_\nu, 
\label{f_lambda_C_mu}
\eea
(where the identity (\ref{C_mu_prod}) has been used in the second line).
\end{lemma}

\subsection{Weighted double Hurwitz numbers: enumerative geometric and combinatorial }

We now proceed to the enumerative geometrical definition of  weighted Hurwitz numbers.
For a fixed pair of branch points, say at $(0, \infty)$,  with ramification profiles $(\mu, \nu)$
and an additional set of  $k$ branch points $(q_1, \dots, q_k)$ with ramification profiles 
$(\mu^{(1)}, \dots , \mu^{(k)})$, we define the
weights to be given by the evaluation of the monomial sum and ``forgotten'' symmetric functions
at the parameter values ${\bf c} =(c_1, c_2, \dots)$ for the two cases corresponding to the
dual weight generating functions $G(z)$ and $\tilde{G}(z)$.
\bea
W_G(\mu^{(1)}, \dots, \mu^{(k)}) &\& := m_\lambda ({\bf c})
\label{W_G_def}
\\
W_{\tilde{G}}(\mu^{(1)}, \dots, \mu^{(k)}) &\& :=
f_\lambda ({\bf c}).
\label{W_G_tilde_def}
\eea

The weighted geometrical Hurwitz numbers $H^d_G(\mu, \nu)$, giving the weighted count
 of such $n$-sheeted branched coverings of the Riemann sphere, having a pair of specified
branch points with ramification profiles  $\mu$ and $\nu$  and any number $k$ of further
branch points, with arbitrary ramification profiles $(\mu^{(1)}, \dots, \mu^{(k)})$,  but fixed genus,
are defined to be the weighted sum
  \bea
H^d_G(\mu, \nu) &\&\deq \sum_{k=0}^\infty \sideset{}{'}\sum_{\substack{\mu^{(1)}, \dots \mu^{(k)} \\ \sum_{i=1}^k \ell^*(\mu^{(i)})= d}}
W_{G}(\mu^{(1)}, \dots, \mu^{(k)}) H(\mu^{(1)}, \dots, \mu^{(k)}, \mu, \nu) 
\label{Hd_G}
\\
H^d_{\tilde{G}}(\mu, \nu) &\&\deq \sum_{k=0}^\infty \sideset{}{'}\sum_{\substack{\mu^{(1)}, \dots \mu^{(k)} \\ \sum_{i=1}^d \ell^*(\mu^{(i)})= d}}
W_{\tilde{G}}(\mu^{(1)}, \dots, \mu^{(k)}) H(\mu^{(1)}, \dots, \mu^{(k)}, \mu, \nu), 
\label{Hd_tildeG}
\eea
where $\sum'$ denotes the sum over all partitions other than the cycle type of the identity element.
The genus $g$ of the covering cover is given by the Riemann-Hurwitz formula \cite{LZ}, 
\be
2 -2g = \ell(\mu) + \ell(\nu) - d.
\label{riemann_hurwitz}
\ee
where
\be
d:= \sum_{i=1}^k \ell^*(\mu^{(i)}).
\ee

The weighted combinatorial Hurwitz numbers $F^d_G(\mu, \nu)$ give weighted enumerations of the paths
 in the Cayley graph of $S_n$ generated by transpositions. 
Expanding the weight generating functions $G(z)$ as  a Taylor series
 \be
 G(z) = 1 + \sum_{i=1}^\infty G_i z^i, \quad \tilde{G}(z) = 1 + \sum_{i=1}^\infty \tilde{G}_i z^i,
 \ee
the weight for a given path depends only upon the signature $\lambda$, and is chosen  to be the product 
of the coefficients of the Taylor series of the generating  functions $G(z)$ (or $\tilde{G}(z))$ corresponding to the parts $\lambda_i$
\bea
G_\lambda := &\&\prod_{i=1}^{\ell(\lambda)} G_{\lambda_i} = e_\lambda({\bf c}) \\
\tilde{G}_\lambda := &\&\prod_{i=1}^{\ell(\lambda)} \tilde{G}_{\lambda_i} =h_\lambda({\bf c}) .
\eea
The path weights for signature $\lambda$, are thus chosen to be either
the products  $e_\lambda({\bf c})$ of the elementary symmetric functions evaluated
at the weighting parameters ${\bf c} =(c_1, c_2, \dots)$ entering in the infinite product representation of $G(z)$ or, 
in the dual case $\tilde{G}(z)$, the products $h_\lambda({\bf c})$ of the complete symmetric functions.

The weighted combinatorial  Hurwitz numbers $F^d_G(z) (\mu, \nu)$ and
$F^d_{\tilde{G}} (\mu, \nu)$  are defined to be the weighted number of $d$-step paths, starting in the
conjugacy class $\cyc(\mu)$ and ending in $\cyc(\nu)$
\bea
F^d_G(\mu, \nu)&\& \deq {1\over n!} \sum_{\lambda, \ \abs{\lambda}=d} e_\lambda({\bf c}) m^\lambda_{\mu \nu}
\label{Fd_G_def}
\\
F^d_{\tilde{G}}(\mu, \nu)&\& \deq  {1\over n!} \sum_{\lambda, \ \abs{\lambda}=d} f_\lambda({\bf c}) m^\lambda_{\mu \nu}
\label{Fd_tilde_G_def}.
\eea

\subsection{Hypergeometric 2D Toda $\tau$-functions as generating functions for weighted Hurwitz number}

Applying the central elements $G_n(z,\JJ) $, $\tilde{G}_n(z,\JJ) $ to the cycle sums $C_\mu$,  $|\mu|=n$,
and using (\ref{G_e_lambda_m_lambda}), (\ref{G_m_lambda_e_lambda}), (\ref{tilde_G_h_lambda_m_lambda}),
( \ref{tilde_G_f_lambda_e_lambda}), (\ref{e_lambda_C_mu}) and  (\ref{f_lambda_C_mu}),  gives \cite{GH2}
\begin{proposition}
\label{G_n_C_mu}
\bea
G_n(z,\JJ) C_\mu &\&=\sum_{d=1}^\infty z^d \sum_{\substack{\nu \\ |\nu|=|\mu|=n}} F^d_G(\mu, \nu) \, C_\nu 
=\sum_{d=1}^\infty z^d \sum_{\substack{\nu \\ |\nu|=|\mu|=n}}H^d_G(\mu, \nu) \, C_\nu \\
\tilde{G}_n(z,\JJ) C_\mu &\&=\sum_{d=1}^\infty z^d \sum_{\substack{\nu \\ |\nu|=|\mu|=n}} F^d_{\tilde{G}}(\mu, \nu) \, C_\nu 
=\sum_{d=1}^\infty z^d \sum_{\substack{\nu \\ |\nu|=|\mu|=n}}H^d_{\tilde{G}}(\mu, \nu) \, C_\nu
\eea
\end{proposition}

This implies, in particular,  that the two definitions of weighted Hurwitz numbers coincide:
\begin{corollary}
\be
F^d_G(\mu, \nu) =  H^d_G(\mu, \nu).
\ee
\end{corollary}

Since
\bea
   G_n (z, \JJ ) F_\lambda &\& = r_\lambda^{G(z)} F_\lambda, \quad |\lambda| =n , \\
   \tilde{G}_n (z, \JJ ) F_\lambda &\& = r_\lambda^{\tilde{G}(z)} F_\lambda,
   \label{content_product_center}
  \eea
the change of basis formulae ({\ref{F_lambda_C_mu}), (\ref{C_mu_F_lambda}) together with
 Proposition~\ref{G_n_C_mu} imply that
\be
\sum_{d=0}^\infty z^d H^d_G(\mu, \nu) = \sum_{d=0}^\infty z^d F^d_G(\mu, \nu)
= \sum_{\substack{\lambda \\ |\lambda|=|\mu|=|\nu|}} r_\lambda^{G(z)}z_\mu^{-1} z_\nu^{-1} \chi_\lambda(\mu) \chi_\lambda(\nu)
\ee
This leads to our main  result  \cite{GH2}
\begin{theorem}
\label{tau_H_G_generating function}
The 2D Toda $\tau$-functions $\tau^{G(z)}({\bf t}, {\bf s})$, $\tau^{\tilde{G}(z)}({\bf t}, {\bf s})$
can be expressed as
\bea
\tau^{G(z)}({\bf t}, {\bf s}) &\&=\sum_{d=0}^\infty z^d \sum_{\substack{\mu, \nu \\ |\mu|=|\nu|}} H^d_G(\mu, \nu) p_\mu({\bf t}) p_\nu({\bf s})
= \sum_{d=0}^\infty z^d \sum_{\substack{\mu, \nu \\ |\mu|=|\nu|}} F^d_G(\mu, \nu) p_\mu({\bf t}) p_\nu({\bf s})
\label{tau_G_H_G}
\\
\tau^{\tilde{G}(z)}({\bf t}, {\bf s}) &\&=\sum_{d=0}^\infty z^d \sum_{\substack{\mu, \nu \\ |\mu|=|\nu|}} H^d_{\tilde{G}}(\mu,  \nu) p_\mu({\bf t}) p_\nu({\bf s})
= \sum_{d=0}^\infty z^d \sum_{\substack{\mu, \nu \\ |\mu|=|\nu|}} F^d_{\tilde{G}}(\mu,  \nu) p_\mu({\bf t}) p_\nu({\bf s}), 
\label{tau_tilde_G_H_G}
\eea
and hence are generating functions for the weighted Hurwitz numbers $H^d_G(\mu, \nu) $, $F^d_G(\mu, \nu)$,
$H^d_{\tilde{G}}(\mu, \nu) $, $F^d_{\tilde{G}}(\mu, \nu)$. 
\end{theorem}

\section{Examples of weighted double Hurwitz numbers}
\label{examples_classical}

We now consider several examples of different types of weighted Hurwitz numbers that are special cases
of this approach.  All these have appeared  in the recent  literature on the subject   \cite{Pa, Ok, GGN1, GGN2, GH1, GH2,  HO3,  AC1,  AC2, H1, H2, Z, KZ, AMMN},  and new examples are easily constructed. Further details for all cases  may be found in \cite{GH1, GH2, HO3, H1, H2}
\subsection{Double Hurwitz numbers for  simple branchings; enumeration of  $d$-step paths in the Cayley graph with equal weight \cite{Pa, Ok}}
\label{ex1}

This was the original case studied by Okounkov  \cite{Ok}, extending an earlier result of Pandharipande \cite{Pa}. 
  The weight generating function in this case is just the exponential $G=\exp$
 \be 
 G(z)= \tilde{G}(z)  = e^z.
 \ee
 The central element $G(z, \JJ) \in \Zb(\Cb[S_n])$ is therefore
 \be
 exp_n(z, \JJ) = e^{z\sum_{b=1}^n \JJ_b}
 \ee
 and the content product formula and fermionic exponent coefficients are given by
\bea
r^{\exp(z)}_j &\&= e^{jz}, \quad  r_\lambda^{\exp} (z)= e^{\frac{z}{2} \sum_{i=1}^{\ell(\lambda)}\lambda_i (\lambda_i - 2i +1)} \\
T^{\exp(z)}_j &\& = {1\over2 } j(j+1)z.
\eea
The generating hypergeometric 2D Toda $\tau$-function is thus
\be
  \tau^{\exp(z)}({\bf t})
   = \sum_{\lambda} e^{\frac{z}{2} \sum_{i=1}^{\ell(\lambda)}\lambda_i (\lambda_i - 2i +1)} s_\lambda ({\bf t}) s_\lambda ({\bf s}),
\ee

   For this case, the infinite product form \eqref{G_inf_prod} of the generating  function  must be interpreted as a limit
\be
e^z =\lim_{m \ra \infty} \left(1+ {z \over m}\right)^m, 
\ee
and the expression (\ref{W_G_def}) for the geometrical weighting becomes:
\be
W_{\exp} (\mu^{(1)}, \dots, \mu^{(k)}) =  \prod_{i=1}^k\delta_{(\ell^*(\mu^{(i)}) , 1)}
 \ee
 (since we require $\ell^*(\mu^{(i)}) \ge 1, \, \forall i$). 
\autoref{tau_H_G_generating function} therefore gives  the generating function
 \be
  \tau^{\exp(z)}({\bf t}) = \sum_{d=0}^\infty z^d \sum_{\substack{\mu, \nu \\ |\mu|=|\nu|}} H^d_{\exp}(\mu,  \nu) p_\mu({\bf t}) p_\nu({\bf s})
 \ee
 function for the (weighted) numbers 
\be
H^d_{\exp}(\mu, \nu) := H(\underbrace{(2, (1)^{n-2}), \dots , (2, (1)^{n-2})}_{d \text{ times }}, \mu, \nu)
\label{H_exp_d_mu_nu}
\ee
of $n$-sheeted branched coverings of the Riemann sphere having $d$ branch points  with simple ramification
 (i.e. , profile $(2, (1)^{n-2})$ and two more (say, at $0$ and $\infty$) with profiles $\mu$ and $\nu$ 
 (weighted, as usual, by the inverse of the automorphism group).

 The combinatorial definition  of the weighted Hurwitz number (\ref{Fd_G_def})  gives 
 \bea
F^d_{\exp}(\mu, \nu) &\&= {1\over n!}\sum_{\lambda, \ \abs{\lambda}=d}{1\over \prod_{i=1}^{\ell(\lambda)}\lambda_i!} m^\lambda_{\mu \nu}\cr
&\&= {1\over d! n!} \times (\#\  d\text{-step paths from an element } h\in\cyc(\mu) \text{ to } \cyc(\nu)).
\eea

\subsection{Coverings with three branch points (Belyi curves): strongly monotonic paths 
 \cite{Z, KZ, AC1, GH2, HO3}}
\label{ex2}
In this case the weight generating function is
\be
G(z) = E(z)\deq 1+ z, 
\ee
so 
\be
c_1 =1, \quad c_i = 0, \quad i>1.
\ee
Therefore the central element $E_n(z, \JJ)\in \Zb(\Cb[S_n])$ is
\bea
E_n(z, \JJ) &\&= \prod_{a=1}^n (1+z \JJ_a), \\
r^{E(z)}_j&\& = 1 + zj, \quad r^{E(z)}_\lambda = \prod_{(i,j)\in \lambda} (1 + z(j-i)) = z^{\abs{\lambda}} \, (1/z)_{\lambda}, \\
T^{E(z)}_j &\& = \sum_{i=1}^j \ln(1+iz), \quad T^{E(z)}_{-j} = -\sum_{i=1}^{j-1}\ln(1-iz), \quad j > 0,
\eea
where
\be
(u)_\lambda := \prod_{i=1}^{\ell(\lambda)}(u-i+1)_{\lambda_i}
\ee
is the multiple Pochhammer symbol corresponding to the partition $\lambda$.

The generating $\tau$-function is thus  \cite{GH1, HO3}
\bea
\tau^{E(z)} ({\bf t}, {\bf s}) &\& = \sum_{\lambda}
z^{\abs{\lambda}} (1/z)_\lambda s_\lambda({\bf t}) s_\mu({\bf s}) \cr
&\& = \sum_{d=0}^\infty z^d \sum_{\mu, \nu,\; \abs{\mu}=\abs{\nu}} H_E^d(\mu, \nu) p_\mu ({\bf t}) p_\nu({\bf s}),
\eea
where
\be
H_E^d(\mu, \nu) = \sideset{}{'}\sum_{\mu^{(1)}, \ \ell^*(\mu_1) =d}
H(\mu^{(1)}, \mu, \nu)
\ee
is  the number of $n=\abs{\mu}=\abs{\nu}=\abs{\mu^{(1)}}$ sheeted branched covers with branch 
points of ramification type $(\mu,\nu)$ at $(0,\infty)$,
and one further branch point, with colength $\ell^*(\mu^{(1)}) =d$; 

These are the double Hurwitz numbers for Belyi curves \cite{Z, KZ, AC1, GH2, HO3}, which enumerate $n$-sheeted branched coverings of the Riemann sphere having three ramification points, with ramification profile  type $\mu$ and $\nu$ at $0$ and $\infty$, and a single additional branch point, with ramification profile $\mu^{(1)}$ of colength
\be
\ell^*(\mu^{(1)}) \deq n - \ell(\mu^{(1)}) = d,
\ee
i.e., with $n-d$ preimages. The genus is again given by the Riemann-Hurwitz formula \eqref{riemann_hurwitz}.

Combinatorially, we have the weight
\be
e_\lambda({\bf c}) = \delta_{\lambda, (1)^{|\lambda|}}
\ee
and therefore
\be
\sum_{\lambda, \abs{\lambda}=d} e_\lambda({\bf c}) m_\lambda(\JJ) = \sum_{b_1 < \dots< b_d} \JJ_{b_1} \cdots \JJ_{b_d}.
\ee
The coefficient $ F_{E}^d(\mu, \nu)$ is thus
\be
F_{E}^d(\mu, \nu) = m^{(1)^{d}}_{\mu \nu},
\label{F_Ed}
\ee
which enumerates all $d$-step paths in the Cayley graph of $S_n$ starting at an element in the conjugacy class of cycle type $\nu$ and ending in the class of type $\mu$, that are strictly monotonically increasing in their second elements \cite{GH2, HO3}.

\subsection{Fixed number of branch points and  genus: multimonotonic paths \cite{HO3} }
\label{ex3}

In this case the weight generating function is :
\be
G(z) =E^k(z)\deq (1+ z)^k, 
\ee
and hence
\be
c_i =1, \ 1\le i \le k, \quad c_i =0, \forall i>k.
\ee
Therefore the central element is
\be
E_n(z, \JJ)^k  = \prod_{a=1}^n (1+z \JJ_a)^k, 
\ee
and
\bea
r^{E^k(z)}_j &\&= (1 + zj)^k, \quad r_\lambda^{E^k(z)} 
= \prod_{(i,j)\in \lambda} (1 + z(j-i))^k = z^{k \abs{\lambda}} ((1/z)_{\lambda})^{k}, \\
T^{E^k(z)}_j &\& = k \sum_{i=1}^j \ln(1+iz), \quad T^{E^k(z)}_{-j} = - k\sum_{i=1}^{j -1}\ln(1-iz), \quad j > 0.
\eea

The generating $\tau$-function is 
\bea
\tau^{E^k (z)} ({\bf t}, {\bf s}) &\& = \sum_{\lambda}
z^{\abs{\lambda}} (1/z)_\lambda s_\lambda({\bf t}) s_\mu({\bf s}) \cr
&\& = \sum_{d=0}^\infty z^d \sum_{\mu, \nu, \ \abs{\mu}=\abs{\nu}} H_{E^k}^d(\mu, \nu) p_\mu ({\bf t}) p_\nu({\bf s}),
\eea
where
\be
H_{E^k}^d(\mu, \nu) = \sideset{}{'}\sum_{\substack{\mu^{(1)}, \dots, \mu^{(k)} \\ \sum_{i=1}^k\ell^*(\mu_i) =d}}
H(\mu^{(1)}, \dots \mu^{(k)}, \mu, \nu)
\ee
is the number of $n=\abs{\mu}=\abs{\nu}=\abs{\mu^{(i)}}$ sheeted branched covers with branch points of ramification  type $(\mu,\nu)$ at $(0,\infty)$,
and (at most) $k$ further branch points, such that the sum of the colengths of their ramification profile  type
(i.e., the ``defect" in the Riemann Hurwitz formula \eqref{riemann_hurwitz}) is equal to $d$:
\be
\sum_{i=1}^k \ell^*(\mu^{(i)}) = kn - \sum_{i=1}^k \ell(\mu^{(i)}) = d.
\ee
This amounts to counting covers with the genus fixed by \eqref{riemann_hurwitz}
and the number of additional branch points fixed at $k$, but no restriction on their simplicity.

The combinatorial weighting for paths of signature $\lambda$ is
\be
e_\lambda({\bf c})= \prod_{i=1}^{\ell(\lambda)}
\binom{k}{\lambda_i} 
\ee
and hence,
\be
 \sum_{\lambda, \abs{\lambda}=d} \left(\prod_{i=1}^{\ell(\lambda)} \binom{k}{\lambda_i}\right) m_\lambda(\JJ) 
= [z^d] \prod_{a=1}^n(1+z J_a)^k
\ee
where $[z^d]$ means the coefficient of $z^d$ in the polynomial.

The weighted combinatorial Hurwitz number
\be
F_{E^k}^d(\mu, \nu) = \sum_{\lambda, \abs{\lambda}=k} \left(\prod_{i=1}^{\ell(\lambda)} \binom{k}{\lambda_i}\right) m^\lambda_{\mu \nu}
\label{F_Ed}
\ee
is thus the number of $(d+1)$-term products $(a_1\, b_1) \cdots (a_d\, b_d) h$ such that $h\in \cyc(\mu)$, while $(a_1\, b_1) \cdots (a_d\, b_d) h \in \cyc(\nu)$, which consist of a product of $k$ consecutive subsequences, each of which is strictly monotonically increasing in their second elements  \cite{GH1, HO3}.

\subsection{Signed Hurwitz numbers at fixed genus: weakly monotonic paths \cite{GGN1, GGN2, GH1}}
\label{ex4}

This case was studied from the combinatorial viewpoint, and related to the HCIZ internal
in  \cite{GGN1, GGN2, GH1}.  It is the dual $\tilde{E}$ of the weight generating function of  \autoref{ex2}.
\be
\tilde{E}(z) := H(z)\deq \frac{1}{1- z} 
\ee
and hence we have
\be
c_i =1, \ 1\le i \le k, \quad c_i =0, \ \forall i>k
\ee
as before, but the relevant combinatorial weighting factor is
\be
 h_\lambda({\bf c}) =1 \quad  \forall \lambda .
 \ee
 The corresponding  central element is
 \be
H_n(z, \JJ) = \prod_{a=1}^n (1-z \JJ_a)^{-1}, 
\ee
and therefore
\bea
r^{H(z)}_j&\& = (1 - zj)^{-1}, \quad r_\lambda^{H(z)} (z) =
\prod_{(i,j)\in \lambda} (1 - z(j-i))^{-1} = (-z)^{-\abs{\lambda}}((-1/z)_\lambda)^{-1}, \\
T^{H(z)}_j &\& = - \sum_{i=1}^j \ln(1-iz), \quad T^{H(z)}_{-j} = \sum_{i=1}^{j -1}\ln(1+iz), \quad j > 0.
\eea

The generating $\tau$-function for this case is \cite{GH1, HO3}
\bea
\tau^{H(z)} ({\bf t}, {\bf s}) &\& = \sum_{\lambda}
(-z)^{-\abs{\lambda}} \left(-1/z\right)^{-1}_{\lambda}
s_\lambda({\bf t}) s_\mu({\bf s}) \cr
&\& = \sum_{d=0}^\infty z^d\sum_{\mu, \nu,\; \abs{\mu}=\abs{\nu}} H^d_{H}(\mu, \nu) p_\mu ({\bf t}) p_\nu({\bf s})
\label{tau_H_s_lambda_exp}
\eea
where
\be
H^d_{H}(\mu, \nu) = (-1)^{n+d}\sum_{k=1}^\infty (-1)^k \sideset{}{'} \sum_{\substack{\mu^{(1)},\dots,\mu^{(k)} \\ \sum_{i=1}^k\ell^*(\mu_i) = d}}
H(\mu^{(1)}, \dots \mu^{(k)}, \mu, \nu)
\ee
is the signed enumeration of $n =\abs{\mu}=\abs{\nu}$ sheeted branched covers with branch points of ramification  type $(\mu,\nu)$ at $(0,\infty)$,
and any number further branch points, the sum of whose colengths is $d $, with sign determined by the parity of the number of branch points \cite{HO3}.

These are thus  double Hurwitz numbers for $n$-sheeted branched coverings of the Riemann sphere with branch points having ramification profile  type $(\mu, \nu)$ at $(0,\infty)$  and an arbitrary number of further branch points,
such that the sum of the colengths of their ramification profile lengths is again equal to $d$
\be
\sum_{i=1}^k \ell^*(\mu^{(i)}) = kn - \sum_{i=1}^k \ell(\mu^{(i)}) = d.
\ee
The latter are counted with a sign, which is $ (-1)^{n+d}$ times the parity of the number of branch points \cite{HO3}. The genus is again given by \eqref{riemann_hurwitz}.

The combinatorial Hurwitz number $F_{H}^d(\mu, \nu)$, derived from
\be
\sum_{\lambda, \abs{\lambda}=d} h_\lambda({\bf c})  m_\lambda(\JJ) = \sum_{b_1 \le \dots \le b_d} \JJ_{b_1} \cdots \JJ_{b_d}.
\ee
is therefore  is given by
\be
F_{H}^d(\mu, \nu) = \sum_{\lambda, \ \abs{\lambda}=k} m^{\lambda}_{\mu \nu},
\ee
which is the number of products of the form $(a_1\, b_1) \cdots (a_d\, b_d) h$ for $g\in \cyc(\mu)$  that are weakly monotonically increasing, such that
$(a_1\, b_1) \cdots (a_d\, b_d) h \in \cyc(\nu)$.
These thus enumerate the $d$-step paths in the Cayley graph of $S_n$ from an element in the conjugacy class of cycle type $\mu$ to the class cycle type $\nu$, that are weakly monotonically increasing in their second elements \cite{GH1}.

Equivalently, they are double Hurwitz numbers for $n$-sheeted branched coverings of the Riemann sphere with branch points at $0$ and $\infty$ having ramification profile  type $\mu$ and $\nu$,
and an arbitrary number of further branch points, such that the sum of the colengths of their ramification profile lengths is again equal to $d$
\be
\sum_{i=1}^k \ell^*(\mu^{(i)}) = kn - \sum_{i=1}^k \ell(\mu^{(i)}) = d.
\ee
The latter are counted with a sign, which is $ (-1)^{n+d}$ times the parity of the number of branch points \cite{HO3}. The genus is again given by \eqref{riemann_hurwitz}.

This case is known to have a matrix model representation \cite{GGN1, GGN2}  when the flow parameters
 ${\bf t}$ and ${\bf s}$ are restricted   to be trace invariants of  a pair of $N\times N$ normal matrices $A$, $B$:
   \be
   t_i = {1\over i} \tr(A^i), \quad  s_i = {1\over i} \tr(B^i),
   \label{t_i_A_B}
   \ee
Within a normalization, setting 
\be
z = -{1\over N}
\label{z_1_over_N}
\ee
as the expansion parameter, we have equality with the HCIZ double matrix integral
\be
\tau^{H(-{1\over N})} ({\bf t}, {\bf s}) =  \II_N(A, B)
    := \int_{\mathrlap{U \in U(N)}} \, e^{ \tr(UAU^\dag B)} d\mu(U)
    = \left(\prod_{k=0}^{N-1} k!\right) \frac{\det\big(e^{a_i b_j}\big)_{1 \leq i,j \leq N}}{\Delta({\bf a}) \Delta ({\bf b})},
    \label{HCIZ}
\ee
where $d\mu(U)$ is the Haar measure on $U(N)$, ${\bf a} = (a_1, \ldots, a_N)$, ${\bf b} = (b_1, \ldots, b_N)$ are the eigenvalues of $A$ and $B$ respectively, and $\Delta({\bf a})$, $\Delta({\bf b})$ are the Vandermonde determinants.

The identification (\ref{z_1_over_N}), however,  gives rise to a cutoff in the expansion (\ref{tau_H_s_lambda_exp}),
giving a sum only over partitions $\lambda$ of length $\ell(\lambda) \le N$:
\be
\tau^{H(-{1\over N})} ({\bf t}, {\bf s})  = \sum_{\lambda, \ \ell(\lambda)\le N}
{N^{|\lambda|} \over (N)_{\lambda} }s_\lambda({\bf t}) s_\mu({\bf s})  .
\ee

\subsection{Quantum weighted branched coverings and paths \cite{GH2}}
\label{ex5}

In \cite{GH2} three variants of quantum Hurwitz numbers were studied, with
weight generating functions denoted $E(q,z)$, $H(q,z)$ and $E'(q,z)$. We only consider the
case $E'(q,z)$, which has the most interesting interpretation in relation to Bosonic gases.
The other two  are developed in detail in \cite{GH2} and may also be obtained as special
cases of the MacDonald polynomial approach to quantum Hurwitz numbers developed in \cite{H2} which is summarized in \autoref{macdonald_quantum_hurwitz} below.

The weight generating function is 
\bea
E'(q,z) &\& \deq \prod_{i=1}^\infty (1+ q^i z) =1 + \sum_{i=0}^\infty E'_i(q) z^i,\\
E'_i(q) &\& \deq \frac{q^{\frac{1}{2}i(i+1)}}{\prod_{j=1}^i (1-q^j)}, \quad i \ge 1,
\eea
where $q$ is viewed as a quantum deformation parameter that may interpreted (see below) in terms 
of the energy distribution of Bosonic gases with a linear  energy spectrum. This is related to the quantum dilogarithm function by
\be
(1+z) E'(q, z) = e^{-\Li_2(q, -z)}, \quad \Li_2(q, z) \deq \sum_{k=1}^\infty \frac{z^k}{k (1- q^k)}.
\ee
We thus have 
\be
c_i = q^i, \quad i \ge 1, \quad 
e_\lambda({\bf c}) = :E'_\lambda(q) = \prod_{i=1}^{\ell(\lambda)}\frac{q^{\frac{1}{2}\lambda_i(\lambda_i +1)}}{\prod_{j=1}^{\lambda_i} (1-q^j)} .
\ee
The  central element $E'_n(q, x\JJ)  \in \Zb(\Cb[S_n])$ is given by
\be
E'_n(q, z\JJ) = \prod_{a=1}^n \prod_{k=1}^\infty (1+q^k z\JJ_a), 
\ee
and hence that content product coefficient is
\bea
r^{E'(q, z)}_j &\&= \prod_{k=1}^\infty (1+ q^k z j), \\
r^{E'(q, z)}_\lambda(z) &\&= \prod_{k=1}^\infty \prod_{(i,j)\in \lambda} (1+ q^k z (j-i))
= \prod_{k=1}^\infty (zq^k)^{\abs{\lambda}} (1/(zq^k))_\lambda.
\eea

The generating $\tau$-function  is therefore  \cite{GH2}
\bea
\tau^{E'(q, z)} ({\bf t}, {\bf s}) &\& = \sum_{\lambda}
\left(\prod_{k=1}^\infty (zq^k)^{\abs{\lambda}} (1/(zq^k))_\lambda \right)
s_\lambda({\bf t}) s_\mu({\bf s}) \cr
&\& = \sum_{d=0}^\infty z^d\sum_{\mu, \nu,\; \abs{\mu}=\abs{\nu}} H^d_{E'(q)}(\mu, \nu) p_\mu ({\bf t}) p_\nu({\bf s})
\label{tau_e_prime_q_s_lambda_exp}
\eea
where
\be
H^d_{E'(q)}(\mu, \nu) \deq \sum_{k=0}^\infty \sideset{}{'}\sum_{\substack{\mu^{(1)}, \dots \mu^{(k)} \\ \sum_{i=1}^k \ell^*(\mu^{(i)})= d}}
W_{E(q)}(\mu^{(1)}, \dots, \mu^{(k)}) H(\mu^{(1)}, \dots, \mu^{(k)}, \mu, \nu)
\label{Hd_Eq}
\ee
is the quantum weighted enumeration of $n =\abs{\mu}=\abs{\nu}$ sheeted branched coverings with genus $g$ given by \eqref{riemann_hurwitz}
and weight $W_{E'(q)}(\mu^{(1)}, \dots, \mu^{(k)})$ for  branched coverings of type $ (\mu^{(1)}, \dots, \mu^{(k)}, \mu, \nu)$ given by
\bea
W_{E'(q)} (\mu^{(1)}, \dots, \mu^{(k)}) &\& \deq {1\over\abs{\aut(\lambda)}}
\sum_{\sigma\in S_k} \sum_{1 \le i_1 < \cdots < i_k}^\infty q^{i_1 \ell^*(\mu^{(\sigma(1))})} \cdots q^{i_k \ell^*(\mu^{(\sigma(k))})} \cr
&\&= {1\over \abs{\aut(\lambda)}}\sum_{\sigma\in S_k} \frac{q^{k \ell^*(\mu^{(\sigma(1))})} \cdots q^{\ell^*(\mu^{(\sigma(k))})}}{
(1- q^{\ell^*(\mu^{(\sigma(1))})}) \cdots (1- q^{\ell^*(\mu^{(\sigma(1))}} \cdots q^{\ell^*(\mu^{(\sigma(k))})})} \cr
&\& ={1\over\abs{\aut(\lambda)}} \sum_{\sigma\in S_k} \frac{1}{
(q^{-\ell^*(\mu^{(\sigma(1))})} -1) \cdots (q^{-\ell^*(\mu^{(\sigma(1))})} \cdots q^{-\ell^*(\mu^{(\sigma(k))})}-1)}, \cr
&\&
\label{W_Eprime_q}
\eea
where $\lambda$ is the partition with parts $\{\ell^*(\mu^{(i)})\}_{i=1. \dots, k}$

The combinatorial Hurwitz number $F_{E'(q)}^d(\mu, \nu)$ giving the weighted enumeration of paths is
\be
F_{E'(q)}^d(\mu, \nu) = \sum_{\lambda, \ \abs{\lambda}=d} \frac{q^{\frac{1}{2}i(i+1)}}{\prod_{j=1}^i (1-q^j)} \, m^\lambda_{\mu \nu}.
\ee
and we have the usual equality
\be
H^d_{E'(q)}(\mu, \nu) = F^d_{E'(q)}(\mu, \nu) .
\ee

\br
{\bf Relation to Bosonic gas distribution.}
If we identify
\be
q \deq e^{-\beta \hbar \omega_0}, \quad \beta = k_B T,
\ee
where $\hbar \omega_0$ is the lowest energy state in a gas of identical Bosonic particles,
 assume the energy spectrum  to consist of integer multiples of $\hbar \omega_0$
\be
\epsilon_k = k \hbar \omega_0,
\ee
and assign the energy
\be
\epsilon(\mu^{(1)}, \dots, \mu^{(k)}) = \sum_{i=1}^k \epsilon_{\ell^*(\mu^{(i)})}
\ee
to a configuration with branching profiles $(\mu^{(1)}, \dots, \mu^{(k)}, \mu, \nu)$,
 the distribution function for Bosonic gases gives the weight
\be
W(\mu^{(1)}, \dots, \mu^{(k)})= \frac{1}{e^{\beta \epsilon(\mu^{(1)}, \dots, \mu^{(k)})}-1}.
\ee
The weighting factor $W_{E'(q)} (\mu^{(1)}, \dots, \mu^{(k)})$  in eq.~\eqref{W_Eprime_q}
 is thus the symmetrized product
\be
W_{E'(q)} (\mu^{(1)}, \dots, \mu^{(k)}) =  {1\over \abs{\aut(\lambda)}}\sum_{\sigma\in S_k}
W(\mu^{(\sigma(1)})   \cdots W(\mu^{\sigma(1)}, \dots, \mu^{\sigma(k)})
\label{W_bosonic_gas_weight}
\ee
of that for each subconfiguration.
\er

In \cite{GH2}, a dual pair of similar weight generating functions $E(q,z)$, $H(q,z)$ were introduced, which correspond to  two slightly different definitions of quantum Hurwitz numbers. These are the $q$-analogs of what, when extended to the Cauchy-Littlewood formula, become the generating functions of the elementary and the complete symmetric functions:
\bea
E(q,z) &\&:= \prod_{k=0}^\infty (1+z q^k) \\
H(q,z) &\& :=  \prod_{k=0}^\infty (1-z q^k)^{-1}
\eea 
The corresponding weights for branched covers with ramification profiles $(\mu^{(1)}, \dots, \mu^{(k)}) $ and 
$ (\nu^{(1)}, \dots, \nu^{(\tilde{k})})$ at the branch points are:
\bea
W_{E(q)} (\mu^{(1)}, \dots, \mu^{(k)}) &\&{1\over \abs{\aut(\lambda)}}
\sum_{\sigma\in S_k} \sum_{0 \le i_1 < \cdots < i_k}^\infty q^{i_1 \ell^*(\mu^{(\sigma(1))})} \cdots q^{i_k \ell^*(\mu^{(\sigma(k))})} \cr
&\&=  {1\over \abs{\aut(\lambda)}}\sum_{\sigma\in S_k} \frac{q^{(k-1) \ell^*(\mu^{(\sigma(1))})} \cdots q^{\ell^*(\mu^{(\sigma(k-1))})}}{
(1- q^{\ell^*(\mu^{(\sigma(1))})}) \cdots (1- q^{\ell^*(\mu^{(\sigma(1))})} \cdots q^{\ell^*(\mu^{(\sigma(k))})})}, \cr
&\&
\label{W_E_q}
\eea
where $\lambda$ is the partition with parts $(\ell^*(\mu^{(1)}), \dots, \ell^*(\mu^{({\tilde{k}})}))$, and
\bea
W_{H(q)} (\nu^{(1)}, \dots, \nu^{(\tilde{k})}) &\& \deq
 {(-1)^{\ell^*{(\lambda)}}\over\abs{\aut(\lambda)}}\sum_{\sigma\in S_{\tilde{k}}} \sum_{0 \le i_1 \le \cdots \le i_{\tilde{k}}}^\infty q^{i_1 \ell^*(\nu^{(\sigma(1))})} \cdots q^{i_k \ell^*(\nu^{(\sigma({\tilde{k}}))})} \cr
&\&=  {(-1)^{\ell^*{(\lambda)}}\over \abs{\aut(\lambda)}}\sum_{\sigma\in S_{\tilde{k}}} \frac{1}{
(1- q^{\ell^*(\nu^{(\sigma(1))})}) \cdots (1- q^{\ell^*(\nu^{(\sigma(1))})} \cdots q^{\ell^*(\nu^{(\sigma({\tilde{k}}))})})}, \cr
&\&
\label{W_H_q}
\eea
where $\lambda$ is the partition with parts $(\ell^*(\nu^{(1)}), \dots, \ell^*(\nu^{({\tilde{k}})}))$. 

The  associated hypergeometric $\tau$-functions  $\tau^{E(q,z))}({\bf t} , {\bf s})$, and $\tau^{H(q,z))}({\bf t} , {\bf s})$ are defined similarly to $\tau^{E'(q,z))}({\bf t} , {\bf s})$ and are generating functions
for the correspondingly modified Hurwitz numbers $F^d_{E(q)}(\mu, \nu) = H^d_{E(q)}(\mu, \nu)$
 and $F^d_{H(q)}(\mu, \nu) = H^d_{H(q)}(\mu, \nu)$. (See \cite{GH2} for further details.)

 For later use, we denote the product of these
\be
W_{Q(q)} (\mu^{(1)}, \dots, \mu^{(k)}; \nu^{(1)}, \dots, \nu^{({\tilde{k}})})
 := W_{E(q)} (\mu^{(1)}, \dots, \mu^{(k)})W_{H(q)} (\nu^{(1)}, \dots, \nu^{({\tilde{k}})}).  
 \label{W_{Q(q)}_mu_nu}
\ee

\section{Multispecies weighted Hurwitz numbers}
\label{multispecies}


\subsection{Hybrid signed Hurwitz numbers at fixed genus: hybrid monotonic paths  \cite{GH1, GH2, HO3}}
\label{ex6.1}

This case is just a hybrid product of the cases of \autoref{ex2} and \autoref{ex4}. 
We choose as generating function
\be
Q(w,z):= {1+ w \over 1-z}
\ee
taking power series in both parameters $(w,z)$.
The associated central element is
\be
Q(w,  z, \JJ) =  E_n(w, \JJ) H_n(z, \JJ) = \prod_{a=1}^n{ 1+w \JJ_a \over 1- z\JJ_a}, 
\ee
and therefore
\bea
r^{Q(w,z)}_j&\& = {1 + jw \over 1-jz} , \\
 r^{Q(w,z)}_\lambda &\&= \prod_{(i,j)\in \lambda} {1 + (j-i)w\over 1-(j-i)z} 
= (-w/z)^{\abs{\lambda}} \, {(1/w)_{\lambda} \over (-1/ z)_\lambda}, \\
T^{Q(w,z)}_j  &\&= \sum_{i=1}^j \ln{1+iw\over 1-iz}, \quad T^{Q(w,z)}_{-j} (w,z)= \sum_{b=1}^{j-1}\ln{1+iz\over 1- iw}, \quad j > 0,
\eea

The generating $\tau$-function is thus  \cite{GH1, HO3}
\bea
\tau^{Q(w, z)} ({\bf t}, {\bf s}) &\& = \sum_{\lambda}
(-w/z)^{\abs{\lambda}}{ (1/w)_\lambda \over (-1/z)_\lambda} s_\lambda({\bf t}) s_\mu({\bf s}) \cr
&\& = \sum_{c=0}^\infty \sum_{d=0}^\infty w^c z^d \sum_{\mu, \nu,\; \abs{\mu}=\abs{\nu}} H^c_d(\mu, \nu) p_\mu ({\bf t}) p_\nu({\bf s}),
\eea
where
\be
H^c_d(\mu, \nu) = \sum_{k=0}^\infty\sum_{\substack{\mu^{(1)} \\ \ell^*(\mu_1) =c}} (-1)^{k+d}
\sideset{}{'}\sum_{\substack{\nu^{(1)},  \dots, \nu^{(k)}\\ \sum_{i=1}^k \ell^*(\nu^{(i)})=d}}
 H(\mu^{(1)}, \nu^{(1)}, \dots , \nu^{(k)}, \mu, \nu)
\ee
is  the number of $n=\abs{\mu}=\abs{\nu}=\abs{\mu^{(1)}}= |\nu^{(1)}|=  \cdots = |\nu^{(k)}|$ sheeted branched 
covers with branch points of ramification  type $(\mu,\nu)$ at $(0,\infty)$,
one further branch point, of ``first class'', with colength $\ell^*(\mu^{(1)}) =c$ and $k$ further branch points, 
$(\nu^{1)}, \dots,  \nu^{(k)})$ of ``second class'' with total colength equal to $d$,
\be
\sum_{i=1}^k \ell^*(\mu^{(i)})=d
\label{k_ell_d_constraint}
\ee
counted with sign $(-1)^{k+d}$ determined by the parity of $k$. Note that the 
sum over $k$ is actually finite, because of the constraint (\ref{k_ell_d_constraint}).
As usual, the Riemann-Hurwitz formula
\be
2-2g = \ell(\mu) +\ell(\nu) -d
\ee
determines the genus $g$ of the covering surface.

The meaning of the combinatorial Hurwitz number $F^c_d(\mu, \nu) = H^c_d(\mu, \nu)$
in this case is clear from combining its meaning for the cases considered in  \autoref{ex2} and \autoref{ex4};
it is the number of $c +d$ step paths  in the Cayley graph of $S_n$ starting at an element $h\in \cyc(\mu)$
 in the conjugacy class of type $\cyc(\mu)$ and ending in $\cyc(\nu)$ such that the first $c$ steps are
 strictly monotonic and the next $d$ steps are weakly monotonic.
 
 This case also has a matrix integral representation, analogous to the HCIZ integral when the flow parameters
 ${\bf t}$ and ${\bf s}$ are again restricted to equal the trace invariants of a pair $A$, $B$ of normal matrices
 as in (\ref{t_i_A_B}), and the expansion parameters are equated to
 \be
 w = {1\over N -\alpha}, \quad  z = -{1\over N}
 \ee
 for some parameter $\alpha$.
 \bea
 \tau^{Q({1 / (N-\alpha}), {-1/ N})} ({\bf t}, {\bf s}) 
  &\& = \int_{\mathrlap{U \in U(N)}} \, \det(\Ib - \zeta UAU^\dag B)^{\alpha-N} d\mu(U)\\
   &\& = \left(\prod_{k=0}^{N-1} {k! \over (1 -\alpha)_k}\right) {\det\left(1- \zeta a_i b_j\right)^{\alpha-1}_{1 \leq i,j \leq N} \over \Delta({\bf a}) \Delta ({\bf b})}
    \label{HCHO}
 \eea
 where
 \be
 \zeta:= {N \over N-\alpha}.
 \ee
 The identification (\ref{z_1_over_N}) again gives rise to a cutoff in the expansion (\ref{tau_H_s_lambda_exp}),
restricting the sum to partitions $\lambda$ of length $\ell(\lambda) \le N$:
\be
 \tau^{Q({1 / (N-\alpha}), {-1/ N})} ({\bf t}, {\bf s})  = \sum_{\lambda, \ \ell(\lambda)\le N}
\left({1 - {\alpha\over N}}\right)^{|\lambda|}  {(N-\alpha)_\lambda\over (N)_{\lambda}} s_\lambda({\bf t}) s_\mu({\bf s})  .
\ee

\subsection{Signed multispecies Hurwitz numbers: hybrid multimonotonic paths \cite{HO3} }
\label{ex6.2}

Now consider the multiparametric generalization of the previous example. We introduce $l +m$
expansion parameters 
\be
{\bf w} := (w_1, \dots, w_l), \quad {\bf z} =(z_1, \dots, z_m).
\ee 
The weight generating functions $G$ is chosen to be products of those for the previous case:
\be
Q^{(l,m)}({\bf w}, {\bf z}) := \prod_{\alpha=1}^l E(w_\alpha)  \prod_{\beta=1}^m H(z_\beta)
= {\prod_{\alpha=1}^l (1+ w_\alpha)\over  \prod_{\beta=1}^m (1 - z_\beta)} 
\ee
The corresponding element  of the  center $\Zb(\Cb[S_n])$ is
\be
Q^{(l,m)}_n({\bf w}, {\bf z}, \JJ) = \prod_{a=1}^nQ^{(l,m)}({\bf w}\JJ_a, {\bf z}\JJ_a), 
\ee 
and therefore the eigenvalues of $Q^{(l,m)}_n({\bf w}, {\bf z}, \JJ)$ are
\be
 r^{Q^{(l,m)}({\bf w}, {\bf z})}_\lambda = \prod_{(i,j)\in \lambda} {\prod_{\alpha=1}^l(1 + (j-i)w_\alpha)\over \prod_{\beta=1}^m (1-(j-i)z_\beta)}  ={\prod_{\alpha=1}^l(w_\alpha)^{|\lambda|}(1/w_\alpha)_{\lambda} \over \prod_{\beta=1}^m(-z_\beta)^{|\lambda| }(-1/ z_\beta)_\lambda}, 
 \ee
 while the diagonal exponential fermionic coefficients are
 \be
T_j^{Q^{(l,m)}({\bf w}, {\bf z})} = \sum_{i=1}^j \ln{\prod_{\alpha=1}^l(1+iw_\alpha)\over \prod_{\beta=1}^m(1-iz_\beta)}, 
\quad T_{-j}^{Q^{(l,m)}({\bf w}, {\bf z})} 
=- \sum_{i=0}^{j-1}\ln{\prod_{\alpha=1}^l(1-iw_\alpha)\over\prod_{\beta=1}^m( 1+ iz_\beta)}, \quad j > 0.
\ee

The generating $\tau$-function is thus  \cite{HO3}
\bea
\tau^{Q^{(l,m)}({\bf w}, {\bf z})} ({\bf t}, {\bf s}) &\& = \sum_{\lambda}
{\prod_{\alpha=1}^l(w_\alpha)^{|\lambda|}(1/w_\alpha)_{\lambda} \over \prod_{\beta=1}^m(-z_\beta)^{|\lambda| }(-1/ z_\beta)_\lambda} s_\lambda({\bf t}) s_\lambda({\bf s}) \cr
&\& = \sum_{{\bf d} \in \Nb^l} \sum_{\tilde{\bf d} \in \Nb^m} {\bf w}^{\bf d} {\bf z}^{\tilde{{\bf d}}}\sum_{\mu, \nu,\; \abs{\mu}=\abs{\nu}} H_{Q^{(l,m)}}^{({\bf d}, \tilde{\bf d})}(\mu, \nu) p_\mu ({\bf t}) p_\nu({\bf s}),
\label{multi_colour_hybrid_signed_tau}
\eea
where multi-index notation has been used:
\be
 {\bf w}^{\bf d} := \prod_{\alpha=1}^l w_\alpha^{d_\alpha}, \quad   {\bf z}^{\tilde{\bf d}}
  := \prod_{\beta=1}^m z_\beta^{\tilde{d}_\beta}
\ee
with
\be
{\bf d} := (d_1, \dots, d_l), \quad \tilde{\bf d} := (\tilde{d}_1, \dots , \tilde{d}_m), \quad d_\alpha, \tilde{d}_\beta \in \Nb.
\ee
Here
\bea
H_{Q^{(l,m)}}^{({\bf d}, \tilde{\bf d})}(\mu, \nu) &\&=  (-1)^{D } \sum_{\{k_\beta\}_{\beta=1}^m} \sum_{\substack{\{\mu^{(\alpha)} \}\\ \ell^*(\mu^{(\alpha)})=d_\alpha}}
\sideset{}{'}\sum_{\substack{\{ \nu^{(\beta, i_\beta)} \}\\\quad  \sum_{i_\beta =1}^{k_\beta}\ell^*(\nu^{(\beta, i_\beta)}) =\tilde{d}_\beta}}
{\hskip -20 pt}(-1)^C H(\{ \mu^{(\alpha)}\}, \{{\nu^{(\beta, i_{\beta})}\}_{i_\beta=1}^{k_\beta}, 
\mu, \nu) },\cr
&\&
 \label{signed_coloured_multihurwitz}
\eea 
is the signed total number of branched coverings, weighted by the inverses of their automorphism groups,
with branch points at $(0, \infty)$ having ramification profiles $(\mu, \nu)$,
and further branch points divided into two types:  $l$ ``plain'' branch points  $\{\mu^{(\alpha)}\}_{\alpha=1, \dots , l}$
with colengths
\be
\ell^*(\mu^\alpha) = d_\alpha
\ee
and
\be
 C= \sum_{\beta=1}^m k_\beta, \quad  
 \ee
``coloured'' branch points with colours labeled by $\beta=1, \dots, m$ and
ramification profiles $\{ \nu^{(\beta,i_\beta)}\}_{\beta=1, \dots, m; \ i_{\beta}=1, \dots k_{\beta}}$, 
of total ramification type colengths
\be
\sum_{i_\beta=1}^{k_{\beta}} \ell^*(\nu^{(\beta, i_{\beta})} )=\tilde{d}_\beta
\ee
in each colour group, and
\be
D = \sum_{\beta=1}^m \tilde{d}_\beta 
\ee
is the sum of these colengths over all colours.
The  genus $g$ of the covering surface is determined by the Riemann-Hurwitz formula:
\be
2-2g =   \ell(\mu) + \ell(\nu) -\sum_{\alpha=1}^l  d_\alpha - D.
\label{riemann_hurwitz_multi_colour_signed}
\ee

The combinatorial significance of the weighted Hurwitz number $F^{({\bf d}, \tilde{\bf d})}_{Q^{(l,m)}}(\mu, \nu)$ in this case is given (see \cite{HO3})  by:
\bt
\label{combinatorial_interpretation}
The coefficients  $ H^{({\bf d}, \tilde{\bf d})}_{Q^{(l,m)}}(\mu, \nu)=F^{({\bf d}, \tilde{\bf d})}_{Q^{(l,m)}}(\mu, \nu)$ in the expansion
 (\ref{multi_colour_hybrid_signed_tau})   are equal to the number of paths in the Cayley graph of  $S_n$ generated by transpositions $(a\, b)$,  $a<b$,  starting at an element in the conjugacy class with cycle type given by the partition $\mu$ and ending in the conjugacy class with cycle type given by partition  $\nu$, such that the paths consist of a sequence of 
\be 
 k:= \sum_{\alpha=1}^l d_\alpha + \sum_{\beta=1}^m \tilde{d}_{\beta}
 \ee
 transpositions $(a_1 b_1) \cdots (a_k b_k)$, divided into $l+m$ subsequences, the first $l$ of which
consist of  $\{d_1, \dots, d_l\}$ transpositions that are strictly monotonically increasing
(i.e.\ $ b_i < b_{i+1}$ for each neighbouring pair of transpositions within the subsequence),
 followed by  $\{\tilde{d}_1, \dots, \tilde{d}_m\}$ subsequences within each of which the
 transpositions are weakly monotonically increasing  (i.e. $b_i \le b_{i+1}$ for each neighbouring pair)
\et

\subsection{General weighted multispecies  Hurwitz numbers \cite{H1, H2} }
\label{ex6.3}

We  may extend the  multispecies  signed Hurwitz numbers  considered  in the preceding section  to general multispecies weighting  \cite{H1, H2} by replacing the factors $E(w_\alpha)$ and $H(z_\beta)$ in the above by arbitrary weight generating functions of type $G^\alpha(w_\alpha)$ and dual type  $\tilde{G}^\beta(z_\beta)$.

The partitions are divided into two classes: those corresponding to the weight factors of type $G(w)$,
 labelled $\{\mu^{(\alpha, u_\alpha)}\}$,  and those corresponding to  dual type $\tilde{G}(z)$,
 labelled $\{\nu^{(\beta, v_\beta}\}$,  These are  further subdivided into $l$ ``colours'', or ``species''  for the first class, 
denoted by the label $\alpha =1, \dots , l$ and $m$ in the second, denoted by $\beta=1, \dots , m$. 
Any given configuration
$\{ \{\mu^{(\alpha, u_\alpha)}\}_{1\le u_\alpha \le k_\alpha}, \{\nu^{(\beta, v_\beta}\}_{1\le v_\beta \le \tilde{k}_\beta}\}$
  has $k_\alpha$ elements of colour $\alpha$  in the first class  and $\tilde{k}_\beta$ elements of colour $\beta$
  in the second class,  for a total of
\be
k = \sum_{\alpha=1}^l k_\alpha  + \sum_{\beta=1}^m \tilde{k}_\beta
\ee
partitions.

Denoting the $l+m$  expansion parameters again as
\be
{\bf w} = (w_1, \dots, w_l), \quad {\bf z} = (z_1, \dots, z_m),
\ee
the multispecies weight generating function is formed from the  product 
\be
G^{(l,m)}({\bf w}, {\bf z}) := \prod_{\alpha=1}^l G^{\alpha}(w_\alpha)  \prod_{\beta=1}^m\tilde{G}^\beta(z_\beta),
\ee
where each  factor has an infinite product representation that is of one of the two types
\bea
G^{\alpha}(w) &\&= \prod_{i=1}^\infty (1 + c_i^\alpha w),  \ \alpha =1, \dots , l\\
\tilde{G}^{\beta}(w) &\&= \prod_{i=1}^\infty (1 - \tilde{c}_i^\beta w),    \ \beta =1, \dots , m.
\eea
for $l+m$ infinite sequences of parameters 
\bea
{\bf c}^\alpha &\&= (c^\alpha_1, c^\alpha_2, \dots), \quad \alpha =1, \dots , l  \\
 \tilde{\bf c}^\alpha  &\&= (\tilde{c}^\alpha_1, \tilde{c}^\alpha_2, \dots),  \quad \beta =1, \dots m.
\eea
The corresponding central element, denoted  
\be
G^{(l,m)}_n({\bf w}, {\bf z}, \JJ) := \prod_{a=1}^n\left(\prod_{\alpha=1}^lG^\alpha(w_\alpha \JJ_a) \right)
\left(\prod_{\beta=1}^m \tilde{G}^\beta(z_\beta \JJ_a)\right)
\ee 
has eigenvalues
\be
r_\lambda^{G^{(l,m)}({\bf w}, {\bf z}) }= \prod_{\alpha=1}^l r_\lambda^{G^\alpha}(w_\alpha) 
  \prod_{\beta=1}^mr_\lambda^{\tilde{G}^\beta}(z_\beta)
 \ee
   in the $\{F_\lambda\}$ basis where, as before,
   \be
   r_\lambda^{G^\alpha}(w_\alpha) := \prod_{(ij)\in \lambda} G(w_\alpha(j-i)),  \quad  r_\lambda^{\tilde{G}^\beta}(z_\beta)
    := \prod_{(ij)\in \lambda} \tilde{G}(z_\beta(j-i)).
   \ee
The diagonal exponential fermionic coefficients are
 \bea
T_j^{G^{(l,m)}({\bf w}, {\bf z})}&\&= \sum_{i=1}^j\left(\prod_{\alpha=1}^lG^\alpha(iw_\alpha)\prod_{\beta=1}^m \tilde{G}^\beta(iz_\beta)\right), \cr
T_{-j}^{G^{(l,m)}({\bf w}, {\bf z})}
&\&=- \sum_{i=0}^{j-1}\ln \left(\prod_{\alpha=1}^lG^\alpha(-iw_\alpha)\prod_{\beta=1}^m \tilde{G}^\beta(-iz_\beta)\right), \quad j > 0.
\eea

The generating hypergeometric $\tau$-function is   \cite{H1}
\bea
\tau^{G^{(l,m)}({\bf w}, {\bf z})}({\bf t}, {\bf s})  &\& = \sum_{\lambda}
 \prod_{\alpha=1}^l r_\lambda^{G^\alpha(w_\alpha)}
  \prod_{\beta=1}^mr_\lambda^{\tilde{G}^\beta((z_\beta)} \,  s_\lambda({\bf t}) s_\lambda({\bf s})  \cr
  &\& = \sum_{{\bf d} \in \Nb^l} \sum_{\tilde{\bf d} \in \Nb^m} {\bf w}^{\bf d} {\bf z}^{\tilde{{\bf d}}}\sum_{\mu, \nu,\; \abs{\mu}=\abs{\nu}} H_{G(l,m)}^{({\bf d}, \tilde{\bf d})}(\mu, \nu) p_\mu ({\bf t}) p_\nu({\bf s}),
\label{multi_colour_hybrid_weighted_tau}
\eea
where
\bea
H_{G^{(l,m)}}^{({\bf d}, \tilde{\bf d})}(\mu, \nu)  
&\&:= \sum_{k_1, \dots, k_l}  \sum_{\tilde{k}_1, \dots , \tilde{k}_m} {\hskip -10 pt} \sum_{\substack{\{\mu^{(\alpha, u_\alpha)}\} \\ |\mu^{(\alpha, u_\alpha)}|=n \\ 
 \sum_{u_\alpha=1}^{k_\alpha} \ell^*(\mu^{(\alpha, u_\alpha)}) = d_\alpha}}
 {\hskip 10 pt}\sideset{}{'}\sum_{\substack{\{\nu^{(\beta, v_\beta)}\} \\ |\nu^{(\beta, u_\beta)}|=n \\ 
 \sum_{v_\beta=1}^{\tilde{k}_\beta} \ell^*(\nu^{(\beta, v_\beta)}) = \tilde{d}_\beta}}\cr
&\&{\hskip 60 pt} \times W_{G^{(l,m)}} (\{\mu^{(\alpha, u_\alpha)}\}, \{\nu^{(\beta, v_\beta}\}) 
H(\{\mu^{(\alpha, u_\alpha)}\}, \{\nu^{(\beta, v_\beta}\}, \mu, \nu)
\cr
&\&
\eea
is the geometrical multispecies Hurwitz number giving the weighted enumeration 
of $n$-sheeted branched coverings with  $l+m$ branch points of type 
$\{ \{\mu^{(\alpha, u_\alpha)}\}_{1\le u_\alpha \le k_\alpha}, \{\nu^{(\beta, v_\beta}\}_{1\le v_\beta \le \tilde{k}_\beta}\}$
and $(\mu, \nu)$ at $(0, \infty)$, with weighting factor  equal to the product of those for single species 
 \bea
W_{G(^{l,m)}} (\{\mu^{(\alpha, u_\alpha)}\}, \{\nu^{(\beta, v_\beta}\}) &\&=  
\prod_{\alpha=1}^l  m_{\lambda^{(\alpha)}} ({\bf c}^{(\alpha)})
\prod_{\beta=1}^m m_{\tilde{\lambda}^{(\beta)}} (\tilde{\bf c}^\beta),
 \eea
 Here  the partitions $\{\lambda^{(\alpha)}\}_{\alpha=1, \dots, l}$,  
 and $\{\tilde{\lambda}^{(\beta)}_{\beta=1, \dots, m}\}$  have lengths
 \be
 \ell(\lambda^{(\alpha)}) = k_\alpha, \quad   \ell(\tilde{\lambda}^{(\beta)}) = \tilde{k}_\beta,
  \ee
  weights
  \be
 |\lambda^{(\alpha)})| = d_\alpha, \quad   |\tilde{\lambda}^{(\beta)}| = \tilde{d}_\beta,
  \ee
 and parts  equal to the colengths $\ell^*(\mu^{(\alpha, u_\alpha)})$ and
$\ell^*(\mu^{(\beta, v_\beta)})$, for $\lambda^{(\alpha)}$ and $\tilde{\lambda}^{(\beta)}$ 
 respectively.
 
  The combinatorial multispecies Hurwitz number $F_{G^{(l,m)}}^{({\bf d}, \tilde{\bf d})}(\mu, \nu) $ is
 determined as follows \cite{H1}. Let $D_n$ denote the number of partitions of weight $n$.
 For each generating function $G^\alpha(w_\alpha)$ or $\tilde{G}^\beta(z_\beta)$, let ${\bf F}_{G^\alpha}^{d_\alpha}$
 and ${\bf F}_{\tilde{G}^\beta}^{\tilde{d}_\beta}$ denote the $D_n \times D_n$ matrices whose elements
 are $F^{d_\alpha}_{G^\alpha}(\mu, \nu)$ and $F^{\tilde{d}_\beta}_{\tilde{G}^\beta}(\mu, \nu)$,
 respectively, as defined in (\ref{Fd_G_def}}), (\ref{Fd_tilde_G_def}). From the fact that the central elements  $\{G^\alpha(w_\alpha, \JJ), \tilde{G}^\beta (z_\beta, \JJ)\}$
all commute, it follows that so do the matrices $\{{\bf F}^{d_\alpha}_{G^\alpha},
{\bf F}^{\tilde{d}_\beta}_{\tilde{G}^\beta}\}$.  Denoting the product of these in any order,
\be
{\bf F}^{({\bf d}, \tilde{\bf d})}_{G^{(l,m)}}:= \prod_{\alpha=1}^l {\bf F}_{G^\alpha}^{d_\alpha} \prod_{\beta=1}^m {\bf F}_{\tilde{G}^\beta}^{\tilde{d}_\beta}, 
\ee
the $(\mu, \nu)$  matrix element $F^{({\bf d}, \tilde{\bf d})}_{G^{(l, m)}}(\mu, \nu)$ is the combinatorial multispecies
weighted Hurwitz number, and is equal to the geometrically defined one.
\be
F^{({\bf d}, \tilde{\bf d})}_{G^{(l, m)}}(\mu, \nu) = H^{({\bf d}, \tilde{\bf d})}_{G^{(l, m)}}(\mu, \nu)
\ee

The combinatorial meaning of $F^{({\bf d}, \tilde{\bf d})}_{G^{(l, m)}}(\mu, \nu)$ is as follows,
Let
\be
d := \sum_{\alpha=1}^l d_\alpha + \sum_{\beta=1}^m \tilde{d}_\beta. 
\ee
Then  $F^{({\bf d}, \tilde{\bf d})}_{G^{(l,m)}}  (\mu, \nu)$, may be interpreted as the weighted sum
over all sequences  of $d$ step paths in the Cayley graph from an element $h \in \cyc(\mu)$ in
the conjugacy class of cycle  type $\mu$ to one $(a_db_d) \cdots (a_1b_1)h$ of cycle  type $\nu$,
 in which the transpositions appearing are subdivided into subsets consisting of $(d_1, \dots, d_l, \tilde{d}_1, \dots , \tilde{d}_m)$  transpositions  in all $ {d! \over (\prod_{\alpha=1}^l d_\alpha !)  (\prod_{\beta=1}^l \tilde{d}_\beta !)}$ possible ways, and to each of these, if the signatures are $(\lambda^{(1)}, \dots, \lambda^{(l)}, \tilde{\lambda}^{(1)} , \dots , \tilde{\lambda}^{(m)})$, a weight is given that is equal to the product
\be
\prod_{\alpha=1}^l e_{\lambda^{(\alpha)}} ({\bf c}^\alpha)  
\prod_{\beta =1}^m h_{\tilde{\lambda}^{(\beta)}} (\tilde{{\bf c}}^\beta) .
\ee
\bt
\be
F^{({\bf d}, \tilde{\bf d})}_{G^{(l, m)}}  (\mu, \nu) = H^{({\bf d}, \tilde{\bf d})}_{G^{(l, m)}}  (\mu, \nu) .
\ee
\et
For proofs of these results, see \cite{H1}.
\section{Quantum weighted Hurwitz numbers  and Macdonald polynomials \cite{H2} }
\label{macdonald_quantum_hurwitz}

Only a summary of the results will be given here; for details  see \cite{H2}.
 
\subsection{Generating functions for Macdonald polynomials}

Following \cite{Mac}, for two infinite sets of indeterminates ${\bf x} := (x_1, x_2, \dots )$, ${\bf y} := (y_1, y_2, \dots )$, we define the generating function
\be
\Pi ({\bf x},  {\bf y}, q, t) := \prod_{a=1}^\infty \prod_{b=1}^\infty  {(t x_a y_b; q)_\infty \over  (x_a y_b; q)_\infty}, 
\ee 
where
\be
(t; q)_\infty := \prod_{i=0}^\infty (1 - t q^i).
\ee
is the infinite $q$-Pochhammer symbol.
$\Pi (q, t, {\bf x},  {\bf y})$  can be expanded in a number of ways in terms of products of dual bases
for the algebra $\Lambda$ of symmetric functions
\bea
\Pi ({\bf x}, {\bf y}, q, t) &\&= \sum_\lambda P_\lambda( {\bf x}, q,t) P_\lambda({\bf y}, q,t)  \\
 &\&= \sum_\lambda z_\mu^{-1}(q,t)p_\lambda({\bf x}) p_\lambda ({\bf y}) \\
 &\&= \sum_\lambda g_\lambda({\bf x}, q,t) m_\lambda ({\bf y})
  \label{Pi_qt_mg}
   \\
 &\&= \sum_\lambda g_\lambda( {\bf y}, q,t) m_\lambda ({\bf x}) 
 \label{Pi_qt_gm}
\eea
 where
 \be
 z_\mu(q,t) := z_\mu n_\mu(q,t), \quad n_\mu :=\prod_{i=1}^{\ell(\mu)} {1 - q^\mu_i \over 1 - t^\mu_i}.
 \ee
Here $\{P_\lambda(q,t, {\bf x})\} $ are the MacDonald symmetric functions, which are orthogonal
\be
 (P_\lambda(q,t), P_\mu(q,t))_{(q,t) }= 0, \quad  \lambda \neq \mu
\ee
 with respect to the inner product $( \ , \  ) _{(q,t)}$ in which the power sum symmetric functions $\{p_\lambda\} $ satisfy
 \be
 (p_\lambda, p_\mu)_{(q,t) } =z^{-1}_\mu (q,t)\delta_{\lambda \mu},
 \ee
$ \{m_\lambda\}$ are the basis of monomial sum symmetric functions and
\be
g_\lambda({\bf x}, q,t)  :=\prod_{i=1}^{\ell(\lambda)}g_{\lambda_i}({\bf x}, q,t) ,  \quad g_j( {\bf x}, q,t)  := (P_j, P_j)^{-1} P_j(q,t, {\bf x}),
\ee
is the $(q,t)$ analog of the interpolating function between the elementary $e_\lambda({\bf x})$ 
and complete $h_\lambda({\bf x})$ symmetric function bases.

\subsection{Quantum families of central elements and  weight generating functions}

We now consider an extended  infinite parametric family of generating functions $M(q, t, {\bf c}, z)$, depending on
the infinite set of ``classical'' parameters ${\bf c}$ appearing in the infinite product representations
 as in (\ref{G_inf_prod}), (\ref{tilde_G_inf_prod})  as well as the further pair of
``quantum deformation'' parameters $(q,t)$ appearing in the  MacDonald polynomials \cite{Mac}.

For a dual pair of ``classical'' generating function $G(z)$, $\tilde{G}(z)$, with 
infinite product representations  (\ref{G_inf_prod}), (\ref{tilde_G_inf_prod})  
we introduce a $(q,t)$ deformed parametric family of weight generating functions $M(q,t, {\bf c}, z)$ as follows
\be
M(q, t, {\bf c}, z) := \prod_{k=0}^\infty G(t q^k z) \tilde{G}( q^k z)
 = \prod_{k=0}^\infty \prod_{i=1}^\infty  {1 - t zq^k c_i \over 1 - zq^k c_i}.
 \label{M_qtcz}
\ee
The associated central element $M_n(q,t, {\bf c}, z,\JJ)\in \Zb(\Cb[S_n])$ is defined as
\be
M_n(q,t, {\bf c}, z \JJ) :=  \prod_{a=1}^b M(q,t, {\bf c}, z\JJ_a) = \Pi({\bf c}, z\JJ, q, t).
\ee
The eigenvectors are the orthogonal idempotents:
\be
M_n(q,t, {\bf c}, z\JJ) F_\lambda = r_\lambda^{M(q,t,{\bf c}, z )} F_\lambda
\ee
and the eigenvalues $r_\lambda^{M(q, t, {\bf c}, z)}$ have  the usual content product form
\be
r_\lambda^{M(q, t, {\bf c}, z)}  =\prod_{(i, j)\in \lambda} M(q,t, {\bf c}, z(j-i)) 
=\prod_{i=1}^\infty {(zt c_i; q)_\lambda \over (zc_i; q)_\lambda}
\label{r_lambda_q_p}
\ee
where,  for a partition $\lambda = (\lambda_ \ge \cdots \ge \lambda_{\ell(\lambda)} >0)$,
the $q$-Pochhammer symbol $(t;q)_\lambda$ is defined as
\be
(t;q)_\lambda := \prod_{i=1}^{\ell(\lambda)} (t q^{-i+1};q)_{\lambda_i}, 
\quad (t;q)_l := \prod_{k=0}^{l-1}(1- tq^k). 
\ee

\subsection{MacDonald  family of quantum weighted Hurwitz numbers}
Using the content product formula (\ref{r_lambda_q_p}), we define the associated
2D Toda $\tau$-function for  $N=0$ as
\be
\tau^{M(q,t, {\bf c}, z)}({\bf t}, {\bf s}):= \sum_{\lambda} r_\lambda^{M(q,t, {\bf c}, z)}  s_\lambda({\bf t}) s_\lambda({\bf s}).
\ee
Substitution of the parameters ${\bf c}$ and the Jucys-Murphy elements $(\JJ_1, \dots , \JJ_n)$ 
for the indeterminates ${\bf x}$ and ${\bf y}$ in  (\ref{Pi_qt_mg}), (\ref{Pi_qt_gm})  gives
 \bea
M_n(q,t, {\bf c}, z \JJ)  &\&= \sum_\lambda g_\lambda(q,t, {\bf c}) m_\lambda ({\bf \JJ})
\label{Mqt_lambda_cJ} \\
 &\&= \sum_\lambda g_\lambda(q,t, {\bf \JJ}) m_\lambda ({\bf c}).
 \label{Mqt_lambda_Jc}
\eea
Applying $M_n(q,t, {\bf c}, z \JJ)$  to $C_\mu$ and using  (\ref{Mqt_lambda_cJ}) and \autoref{m_lambda_JJ_C_mu},
gives
\be
M_n(q,t, {\bf c}, z\JJ) C_\mu = \sum_{d=0}^\infty z^d\sum_{\nu, |\nu|=|\mu|} F^d_{M(q,t,{\bf c})} (\mu, \nu) z_\nu C_\nu
\label{Pi_Cmu_F_Gd}
\ee
where we define, as before, the combinatorial quantum  weighted Hurwitz numbers  $F^d_{M(q,t, {\bf c})}(\mu, \nu)$
associated to the weight generating function $M(q,t, {\bf c}, z)$ as
\be
F^d_{M(q,t,{\bf c})}(\mu, \nu) \deq {1 \over \abs{n}!} \sum_{\lambda, \ \abs{\lambda}=d} g_\lambda({\bf c},q,t) m^\lambda_{\mu \nu}.
\label{Fd_Mqt_def}
\ee

It follows as before that when the $\tau$-function $\tau^{M(q,t, {\bf c}, z)}({\bf t}, {\bf s})$ is
expanded in the basis of products of power sum symmetric functions, the coefficients
are the quantum weighted Hurwitz combinatorial numbers $F^d_{M(q,t,{\bf c})}(\mu, \nu) $:
\be
\tau^{M(q,t, {\bf c}, z)} ({\bf t}, {\bf s})
= \sum_{d=0}^\infty \sum_{\substack{\mu, \nu \\ \abs{\mu} =
 \abs{\nu}}} z^d F^d_{M(q,t,{\bf c})}(\mu, \nu) p_\mu({\bf t}) p_\nu({\bf s}).
\label{tau_GM_F}
\ee
  
The corresponding geometrically defined quantum weighted Hurwitz numbers are somewhat
more intricate.  Let $\{\{\mu^{i, u_i}\}_{u_i =1, \dots , k_i}, \{\nu^{i, v_i}\}_{v_ i= 1, \dots, \tilde{k}_i}, \mu, \nu \}_{i=1, \dots, l}$ 
 denote the branching profiles of an $n$-sheeted covering, of the Riemann sphere,   
with two specified branch points of ramification profile types $(\mu, \nu)$, at $(0, \infty)$,  and the rest divided into  two classes I and II,
denoted  $\{\mu^{(i,u_i)}\}_{u_i =1, \dots , k_i}$ and  $\{\nu^{(i, v_ i)}\}_{v_ i= 1, \dots, \tilde{k}_i}$, respectively. These are further subdivided into $l $ species, or ``colours'', labelled by  $i=1, \dots  l$,  the elements within each colour group distinguished  by the  labels $(u_i =1, \dots , k_i)$  
  and $(v_ i =1, \dots , \tilde{k}_ i)$.  To such a grouping, we assign a partition $\lambda$ of length
\be
\ell(\lambda) =: l
\ee
and weight
\be
d := |\lambda| = \sum_{i =1}^l \left (\sum_{u_i =1}^{k_i}\ell^*(\mu^{(i, u_i)})
+ \sum_{v_i =1}^{\tilde{k}_i}\ell^*(\nu^{(i, v_i)}) \right) = \sum_{i=1}^l d_i,
\ee
whose parts $(\lambda_1\ge \cdots \ge \lambda_l > 0)$ are equal  the total colengths
\be
d_i := \sum_{u_i =1}^{k_i} \ell^*(\mu^{(i, u_i)}) + \sum_{v_i=1}^{\tilde{k}_i}\ell^*(\nu^{(i, v_i)})
\ee
in weakly decreasing order.
By the Riemann-Hurwitz formula, the genus $g$ of the covering curve is given by
\be
  2-2g = \ell(\mu) +\ell(\nu)  - d.
  \ee 
  
We now assign a weight $W_{Q(q)} (\{\mu^{(i,u_i)}, \nu^{( i, v_ i)}\}, {\bf c} ) $ 
to each such covering as in (\ref{W_{Q(q)}_mu_nu}), consisting of the product of all the weights 
$W_{E(q)}(\{\mu^{(i, u_i)}\}_{u_i = 1, \dots, k_i})$, 
$W_{H(q)}(\{\nu^{( i, v_ i)}\}_{v_ i = 1, \dots, \, \tilde{k}_ i})$
for the subsets of different colour and class with the weight $m_\lambda({\bf c})$
given by the monomial symmetric functions evaluated at the parameters ${\bf c}$
\be
  W_{Q(q)} (\{\mu^{(i,u_i)}, \nu^{( i, v_ i)}\}, {\bf c} ) 
 := W_{Q(q)} (\{\mu^{(i,u_i)}, \nu^{( i, v_ i)}\} )m_\lambda({\bf c})
 \ee
 where
 \be
  W_{Q(q)} (\{\mu^{(i,u_i)}, \nu^{( i, v_ i)}\} ) := 
  \prod_{i=1}^lW_{E(q)}(\{\mu^{(i,u_i)}\}_{u_i = 1, \dots, \, k_i})
   W_{H(q)}(\{\nu^{( i, v_ i)}\}_{ i = 1, \dots, \,  \tilde{k}_ i})
   \ee

Using these weights, for every pair $(d,e)$ of non-negative integers and  $(\mu, \nu)$
of partitions of $n$,  we define the  geometrical  quantum weighted Hurwitz numbers 
$H^{(d,e)}_{({\bf c}, q)}(\mu, \nu) $ as the sum
\be
H^{(d,e)}_{({\bf c}, q)}(\mu, \nu) := z_\nu \sum_{l=0}^{d} {\hskip -35 pt}
\sideset{}{'} \sum_{\substack{\{\mu^{(i, u_i)}, \nu^{(i, v_i)}\} , \ k_i\ge 1, \ \tilde{k}_i \ge 1\\
\sum_{i=1}^l \sum_{u_i =1}^{k_i}\ell^*(\mu^{(i,u_i)}) = e, \\  
\sum_{i=1}^l\left( \sum_{u_i =1}^{k_i}\ell^*(\mu^{(i, u_i)} )
+  \sum_{v_ i =1}^{\tilde{k}_ i}\ell^*(\nu^{( i, v_ i)})\right) =d}} {\hskip - 20 pt}
{\hskip-50 pt}W_{Q(q)} (\{\mu^{(i, u_i)}, \nu^{( i, v_ i})\}, {\bf c}) \ 
H(\{\mu^{(i, u_i)}\}_{\substack{u_i = 1, \dots, k_i \\ i =1, \dots , l}} ,
\{\nu^{( i, v_ i)}\}_{\substack{v_ i = 1, \dots, \tilde{k}_ i \\  i =1, \dots , l}},  \mu, \nu).
\label{Hde_c_q_mu_nu}
\ee

We then have the following theorem, which is proved in \cite{H2}:

\begin{theorem}
\label{geometric_hurwitz}
The combinatorial Hurwitz numbers $F^d_{M(q,t,{\bf c})} (\mu, \nu) $  are degree $d$  polynomials in $t$,  whose 
coefficients are equal to  the  geometrical  quantum weighted Hurwitz numbers $H^{(d,e)}_{({\bf c}, q)}(\mu, \nu)$
\be
F^d_{M(q,t,{\bf c})} (\mu, \nu) = \sum_{e=0}^d H^{(d,e)}_{({\bf c}, q)}(\mu, \nu) t^e.
\label{Fd_G_Hde_G} 
\ee

Hence $\tau^{M(q,t,{\bf c}, z)} ({\bf t}, {\bf s})$,  when expanded in
the basis of products of power sum symmetric functions and power series in $z$ and $t$
is the generating function for the  $H^{(d,e)}_{({\bf c}, q)}(\mu, \nu)$'s:
\be
\tau^{M(q,t,{\bf c}, z)} ({\bf t}, {\bf s}) = \sum_{d=0}^\infty \sum_{e=0}^d z^d t^e H^{(d,e)}_{({\bf c}, q)}(\mu, \nu) p_\mu({\bf t}) p_\nu({\bf s}).
\label{tau_G_H}
\ee
\end{theorem}

\subsection{Examples}
We  now give several examples of special classes of weighted Hurwitz numbers
that arise  through restrictions or limits involving the parameters $(q,t,z)$. The details for all these examples are provided in \cite{H2}; we  limit ourselves to specifying the restrictions and limits involved, giving only the generating functions, $\tau$-functions and quantum weighted Hurwirz  formulae for each case.

\subsubsection{Elementary quantum weighted Hurwitz numbers}
In this case, we take the limits $z\ra 0$, $t\ra \infty$, but keeping the value of $-tz$ fixed
at a finite value, that is renamed $z$. The resulting weight generating function is 
\be
E(q, {\bf c}, z)  := \prod_{k=0}^\infty \prod_{i=1}^\infty(1 +zq^k c_i) = \prod_{i=1}^\infty (-zc_i; q)_{\infty}.
 =:\sum_{j=0}^\infty e_j(q,{\bf c}) z^j
\ee
Here $e_j(q, {\bf c})$ is the quantum deformation of the elementary symmetric function $e_j({\bf c})$. 
The corresponding central element $E_n(q, {\bf c}, z, \JJ)  \in \Zb(\Cb[S_n])$ is:
\be
E_n(q, {\bf c}, z \JJ)  := \prod_{a=1}^n E(q, {\bf c}, z\JJ_a) 
 =\sum_{\lambda} z^{|\lambda|} e_\lambda(q, {\bf c})  m_\lambda(\JJ) 
 = \sum_{\lambda} z^{|\lambda|} m_\lambda(\JJ)  e_\lambda(q, {\bf c}) , 
\ee
where 
\be
e_\lambda(q, {\bf c}) := \prod_{i=1}^{\ell(\lambda)} e_{\lambda_i} ({\bf c}).
\ee

Applying $E_n(q, {\bf c}, z\JJ)$ to the orthogonal idempotents
$\{F_\lambda\}$ to obtain the content product coefficients
and to  the cycle sums $\{C_\mu\}$  to obtain the Hurwitz numbers, 
the resulting hypergeometric $2D$ Toda $\tau$-function is 
\bea
\tau^{E(q, {\bf c}, z)}({\bf t}, {\bf s}) &\&= \sum_\lambda r_\lambda^{E(q,{\bf c}, z)} s_\lambda({\bf t}) s_\lambda({\bf s}) \\
 &\&= \sum_{d=0}^\infty  z^d\sum_\lambda F^d_{E(q, {\bf c})}(\mu, \nu)  p_\mu({\bf t}) p_\nu({\bf s}),
\eea
where
\be
r_\lambda^{E(q, {\bf c}, z)} := \prod_{(ij) \in \lambda} \prod_{k=0}^\infty   (-zc_k; q)_\infty
\ee
is the content product coefficient and
\be
F^d_{E(q, {\bf c})}(\mu, \nu) := \sum_{|\lambda|=d}e_\lambda(q, {\bf c}) m^\lambda_{\mu \nu}
\label{F_dE_qc}
\ee
is  the weighted number of $d$-step paths in the Cayley graph of $S_n$ generated by transpositions,
starting at  the conjugacy class $\cyc(\mu)$ and ending at $\cyc(\nu)$, with
weight  $e_\lambda(q, {\bf c})$  for a path of signature $\lambda$. 
  
 Now consider again $n$-fold branched coverings of $\Cb \Pb^1$ with a fixed pair of branch points   at $(0, \infty)$
   with ramification profiles $(\mu, \nu)$ and  a further $  \sum_{i=1}^l k_i $  branch points  $\{\mu^{(i,u_i)}\}_{u_i = 1, \dots, \, k_i}$ of $l$ 
   different species (or ``colours''), labelled by $i=1, \dots , l$, with non trivial ramifications profiles.
   The weight  $W_{E^l(q)} (\{\mu^{(i,u_i)}\}_{\substack{u_i = 1, \dots, k_i \\ i=1, \dots , l}}, {\bf c} )$
for such a covering is defined by
\be
  W_{E^l(q)} (\{\mu^{(i,u_i)}\}_{\substack{u_i = 1, \dots, k_i \\ i=1, \dots , l}},  {\bf c}  )
:= W_{E^l(q)} (\{\mu^{(i,u_i)}\}_{\substack{u_i = 1, \dots, k_i \\ i=1, \dots , l}}  )\, m_\lambda({\bf c})
  \ee
  where
  \be
  W_{E^l(q)}(\{\mu^{(i,u_i)}\}_{\substack{u_i = 1, \dots, k_i \\ i=1, \dots , l}} ) := 
  \prod_{i=1}^lW_{E(q)}(\{\mu^{(i,u_i)}\}_{u_i = 1, \dots, \, k_i}).
\ee

It follows from the general result that 
\be
 F^d_{E(q,{\bf c})}(\mu, \nu) =H^d_{E(q, {\bf c})} (\mu, \nu), 
\ee
where
\be
H^d_{E({\bf c}, q)}(\mu, \nu) := z_\nu \sum_{l=0}^{d} {\hskip -10pt}
\sideset{}{'} \sum_{\substack{\{\mu^{(i, u_i)}\} , \ k_i\ge 1, \ \\
\sum_{i=1}^l \sum_{u_i =1}^{k_i}\ell^*(\mu^{(i,u_i)}) = d 
}} 
{\hskip-20 pt}W_{E^l(q)} (\{\mu^{(i, u_i)}\}_{\substack{u_i = 1, \dots, k_i \\ i=1, \dots , l}}, {\bf c}) \ 
H(\{\mu^{(i, u_i)}\}_{\substack{u_i = 1, \dots, k_i \\ i =1, \dots , l}} ,
 \mu, \nu)
\label{Eq_d_c}
\ee
 is the geometrically defined quantum weighted Hurwitz number for this case.
  
  
\subsubsection{Complete quantum weighted Hurwitz numbers}

This is the dual of the preceding case,  obtained by setting $t=0$. The weight generating function
becomes
\be
H(q, {\bf c}, z) := \prod_{k=0}^\infty \prod_{i=1}^\infty(1 - zq^k c_i)^{-1} = \prod_{i=1}^\infty (zc_i; q)^{-1}_{\infty}
 =:\sum_{j=0}^\infty h_j(q,{\bf c}) z^j,
\ee
where $h_j(q, {\bf c})$ is the quantum deformation of the complete symmetric function $h_j({\bf c})$.  
The corresponding central element $H_n(q, {\bf c}, z, \JJ)  \in \Zb(\Cb[S_n])$  is:
\be
H_n(q, {\bf c}, z \JJ)  := \prod_{a=1}^n H(q, {\bf c}, \JJ_a) 
=\sum_{\lambda}z^{|\lambda|}  h_\lambda(q, {\bf c} )m_\lambda(\JJ) 
 = \sum_{\lambda} z^{|\lambda|} m_\lambda({\bf c})  h_\lambda(q, \JJ) ,
\ee
where 
\be
h_\lambda(q, {\bf c}) := \prod_{i=1}^{\ell(\lambda)} h_{\lambda_i} (q, {\bf c}).
\ee

By specializing the general case by setting $t=0$, the resulting hypergeometric $2D$ Toda $\tau$-function is
\bea
\tau^{H(q,{\bf c}, z)}({\bf t}, {\bf s}) &\&= \sum_\lambda r_\lambda^{H(q, {\bf c}, z)} s_\lambda({\bf t}) s_\lambda({\bf s}) \\
 &\&= \sum_{d=0}^\infty  z^d\sum_\lambda F^d_{H(q,{\bf c})}(\mu, \nu)  p_\mu({\bf t}) p_\nu({\bf s}), 
\eea
where
\be
r_\lambda^{H(q,{\bf c}, z)} := \prod_{(ij) \in \lambda} \prod_{k=0}^\infty   (z(j-i)c_k; q)^{-1}_\infty
\ee
and
\be
F^d_{H(q,{\bf c})}(\mu, \nu) := \sum_{|\lambda|=d}h_\lambda(q, {\bf c}) m^\lambda_{\mu \nu}
\label{F_dH_qc}
\ee
is  the weighted number of paths in the Cayley graph of $S_n$ generated by transpositions,
starting at  the conjugacy class $\cyc(\mu)$ and ending at $\cyc(\nu)$, with 
weight  $h_\lambda(q, {\bf c})$  for a path of signature $\lambda$. 

Consider  again the $n$-fold branched coverings of $\Cb \Pb^1$, with a fixed pair of branch points   at $(0, \infty)$
   with ramification profiles $(\mu, \nu)$ and  a further 
   $  \sum_{i=1}^l \tilde{k}_i $  branch points  $\{\nu^{(i,v_i)}\}_{v_i = 1, \dots, \, \tilde{k}_i}$
   of $l$  different species (or ``colours''), labelled by $i=1, \dots , l$, with nontrivial ramifications profiles.
   The weight  $W_{H^l(q)} (\{\nu^{(i,v_i)}\}_{\substack{v_i = 1, \dots, \tilde{k}_i \\ i=1, \dots , l}}, {\bf c} )$
for such a covering is defined by
\bea
  W_{H^l(q)} (\{\nu^{(i,v_i)}\}_{\substack{v_i = 1, \dots, \tilde{k}_i \\ i=1, \dots , l}},  {\bf c}  )
  &\&:= W_{H^l(q)} (\{\nu^{(i,v_i)}\}_{\substack{v_i = 1, \dots, \tilde{k}_i \\ i=1, \dots , l}}  ) \, m_\lambda({\bf c})\\
  W_{H^l(q)}(\{\nu^{(i,v_i)}\}_{\substack{v_i = 1, \dots, k_i \\ i=1, \dots , l}} ) &\&:= 
  \prod_{i=1}^lW_{H(q)}(\{\nu^{(i,v_i)}\}_{v_i = 1, \dots, \, k_i}).
   \eea
We again have
\be
 F^d_{H(q,{\bf c})}(\mu, \nu) =H^d_{H(q, {\bf c})} (\mu, \nu), 
\ee
where
\be
H^d_{H({\bf c}, q)}(\mu, \nu) := z_\nu \sum_{l=0}^{d} {\hskip -10pt}
\sideset{}{'} \sum_{\substack{\{\nu^{(i, v_i)}\} , \ \tilde{k}_i\ge 1, \ \\
\sum_{i=1}^l \sum_{v_i =1}^{k_i}\ell^*(\nu^{(i,v_i)}) = d 
}} 
{\hskip-20 pt}W_{H^l(q)} (\{\nu^{(i, v_i)}\}_{\substack{v_i = 1, \dots, \tilde{k}_i \\ i=1, \dots , l}}, {\bf c}) \ 
H(\{\nu^{(i, v_i)}\}_{\substack{v_i = 1, \dots, \tilde{k}_i \\ i =1, \dots , l}} ,
 \mu, \nu).
\label{Eq_d_c}
\ee
  
\subsubsection{Hall-Littlewood function weighted Hurwitz numbers}

The generating function for Hall-Littlewood polynomials $P_\lambda({\bf x}, t)$, is obtained
by setting $q=0$ in eq.~(\ref{M_qtcz}). The orthogonality relations \cite{Mac} become
\be
(P_\lambda, P_\mu)_t= \delta_{\lambda \mu} (b_\lambda(t))^{-1},  \quad  b_\lambda (t) := \prod_{i\ge 1} \prod_{k=1}^{m_i(\lambda)} (1 - t^k)
\ee
 with respect to the scalar product $(\  , \ )_t$ defined by
 \be
 (p_\lambda, p_\mu)_t =  \delta_{\lambda \mu} z_\lambda n_\lambda (t), \quad n_\lambda:= \prod_{i=1}^{\ell(\lambda)}  {1\over 1 - t^{\lambda_i}}.
 \ee
 
Defining, as in \cite{Mac}, 
 \be
 q_\lambda({\bf x}, t) := b_\lambda(t) \prod_{i=1}^{\ell(\lambda)} P_j({\bf x}, t)
 \label{q_lambda}
 \ee
we obtain the following expansion
 \be
L(t, {\bf x}, {\bf y})= \prod_{i, j}^\infty {1 - t x_i y_j   \over 1- x_i y_j}
=  \sum_{\lambda}^\infty  q_\lambda({\bf x}, t) m_\lambda({\bf y}) 
= \sum_{\lambda}^\infty   q_\lambda({\bf y}, t) m_\lambda({\bf x}).
\label{Lt_yx_t}
\ee
The corresponding central element of $\Cb[S_n]$ is then
\be
L(t, {\bf c}, z\JJ):=   \prod_{i=1}^\infty \prod_{a=1}^n  {1 - t c_i z \JJ_a  \over 1- c_i z \JJ_a} 
=  \sum_{\lambda}^\infty z^{|\lambda|}q_\lambda({\bf c}, t) m_\lambda({\bf \JJ}) \\
 = \sum_{\lambda}^\infty z^{|\lambda|} q_\lambda({\bf \JJ}, t) m_\lambda({\bf c}).
\label{HL_generating_element}
\ee

The  hypergeometric $2D$ Toda $\tau$-function then reduces to
\bea
\tau^{L(t,{\bf c}, z)}({\bf t}, {\bf s}) &\&:= \sum_\lambda r_\lambda^{L(t,{\bf c}, z)} s_\lambda({\bf t}) s_\lambda({\bf s}) \\
 &\&= \sum_{d=0}^\infty  z^d\sum_\lambda F^d_{L(t,{\bf c})}(\mu, \nu)  p_\mu({\bf t}) p_\nu({\bf s})\\
\eea
where the content product coefficient $r_\lambda^{L(t,{\bf c}, z)}$
is
\be
r_\lambda^{L(t,{\bf c}, z)} := \prod_{(ij) \in \lambda} \prod_{k=1}^\infty  { 1 - t z (j-i) c_k \over 1 - z (j-i) c_k} 
= \prod_{k=1}^\infty (-t)^{|\lambda|} {(- 1/(tz c_k))_\lambda \over (- 1/(z c_k))_\lambda }, 
\ee
and
\be
F^d_{L(t,{\bf c})}(\mu, \nu) := \sum_{|\lambda|=d}q_\lambda({\bf c}, t) m^\lambda_{\mu \nu}
\ee
is  the weighted number of paths in the Cayley graph with the
weight  $q_\lambda({\bf c},t)$  for a path of signature $\lambda$.

As in the general case, we also have
\be
 F^d_{L(t,{\bf c})}(\mu, \nu) = \sum_{e=0}^d H^{(d,e)}_{{\bf c}} (\mu, \nu) t^e
\ee
where, denoting  the total number of branch points, 
\be
K:= \sum_{i=1}^l (k_i +\tilde{k}_i)
\ee
the weighted generalization of the multispecies hybrid signed Hurwitz numbers studied in \cite{HO2} is
\be
H^{(d,e)}_{L({\bf c})}(\mu, \nu) := z_\nu \sum_{l=0}^{d} {\hskip -35 pt}
\sideset{}{'} \sum_{\substack{\{\mu^{(i, u_i)}, \nu^{(i, v_i)}\} , \ k_i\ge 1, \ \tilde{k}_i \ge 1\\
\sum_{i=1}^l \sum_{u_i =1}^{k_i}\ell^*(\mu^{(i,u_i)}) = e, \\  
\sum_{i=1}^l\left( \sum_{u_i =1}^{k_i}\ell^*(\mu^{(i, u_i)} )
+  \sum_{v_ i =1}^{\tilde{k}_ i}\ell^*(\nu^{( i, v_ i)})\right) =d}} {\hskip - 20 pt}
{\hskip-50 pt} (-1)^{K+d-e}
H(\{\mu^{(i, u_i)}\}_{\substack{u_i = 1, \dots, k_i \\ i =1, \dots , l}} ,
\{\nu^{( i, v_ i)}\}_{\substack{v_ i = 1, \dots, \tilde{k}_ i \\  i =1, \dots , l}},  \mu, \nu).
\label{H_de_c}
\ee

Its interpretation in terms of weighted enumerations of multispecies Hurwitz numbers of two classes
is the same as in the general Macdonald case, with the general quantum weighting factor
reducing to a sign times the standard classical one $m_\lambda({\bf c})$.
  

\subsubsection{Jack function weighted Hurwitz numbers}

The Jack polynomials $P^{(\alpha)}_\lambda$ are obtained by setting $t=q^\alpha$ and taking the limit $q \ra 1$.
These satisfy the orthogonality relations \cite{Mac}
\be
\langle P^{\alpha}_\lambda, P^{\alpha}_\mu\rangle_\alpha= \delta_{\lambda \mu} z_\lambda b_\lambda(t),  \quad  b_\lambda (t) := \prod_{i=1}^{\ell(\lambda)} \prod_{j=1}^{\lambda_i} (1 - t^j)
\ee
where the scalar product $\langle \  , \ \rangle_\alpha$ is defined by \cite{Mac}
 \be
 \langle  p_\lambda, p_\mu\rangle_\alpha =  \delta_{\lambda \mu} z_\lambda \alpha^{\ell(\lambda)}. 
 \ee
 
The corresponding family of weight generating functions becomes
\be
J(\alpha, {\bf c}, z) :=\prod_{k=1}^\infty (1 - z c_i)^{-1/\alpha}
\ee
and the central elements
\be
J(\alpha, z \JJ):= \prod_{i=1}^\infty \prod_{a=1}^n(1 - z c_i \JJ_a)^{-1/\alpha} 
 = \sum_{\lambda} z^{|\lambda|} g^{\alpha}_\lambda(\JJ) m_\lambda({\bf c}) 
  = \sum_{\lambda} z^{|\lambda|} g^{\alpha}_\lambda({\bf c}) m_\lambda({\JJ}) ,
\ee
where the symmetric functions $g^{\alpha}_\lambda({\bf x})$ are the analogs of  the $e_\lambda({\bf x})$ or $h_\lambda ({\bf x})$ bases formed from products of  elementary or  complete symmetric functions \cite{Mac}
\be
g^{\alpha}_\lambda({\bf x}) = \alpha^{\ell(\lambda)}\prod_{i=1}^{\ell(\lambda} P^{(\alpha)}_{(\lambda_i)}({\bf x}).
\ee

The associated hypergeometric $2D$ Toda $\tau$-function is
\be
\tau^{J(\alpha, {\bf c}, z)}({\bf t}, {\bf s}) = \sum_{\lambda} r_\lambda^{J(\alpha, {\bf c}, z)} s_\lambda({\bf t}) s_\lambda({\bf s})
\ee
where the content product coefficients are
\be
r_\lambda^{J(\alpha, {\bf c}, z)}  := \prod_{(ij)\in \lambda} \prod_{k=0}^\infty (1- z (j-i) c_k)^{-1/\alpha}
= \prod_{k=1}^\infty (1-zc_k)^{|\lambda|\over \alpha} (-1/z c_k)_\lambda^{-1/\alpha}.
\ee
Expanding over products of power sum  symmetric functions gives
\be
\tau^{J(\alpha, {\bf c}, z)} ({\bf t}, {\bf s})
= \sum_{d=0}^\infty \sum_{\substack{\mu, \nu \\ \abs{\mu} =
 \abs{\nu}=n}} z^d F^d_{J(\alpha, {\bf c})}(\mu, \nu) p_\mu({\bf t}) p_\nu({\bf s})
\label{tau_GJ_F}
\ee
where
\be
F^d_{J(\alpha, {\bf c})} (\mu, \nu) = \sum_{\lambda} g^{\alpha}_\lambda({\bf c})m^\lambda_{\mu \nu}
\ee
is the combinatorial Hurwitz number giving the weighted number of  $d$-step paths of signature $\lambda$ in the Cayley
graph of $S_n$, starting in the conjugacy class $\cyc(\mu)$ and ending in $\cyc(\nu)$, with weight $g^{\alpha)}_\lambda({\bf c})$.

We again have equality with the weighted  geometrical Hurwitz number 
\be
F^d_{J(\alpha, {\bf c})} (\mu, \nu)  = H^d_{J(\alpha, {\bf c})} (\mu, \nu) ,
\ee
where
\be
H^d_{J(\alpha, {\bf c})} (\mu, \nu)  :=  \sum_{k=0}^\infty \left({-{1\over \alpha} \atop k}\right)
  \sum_{\substack{\mu^{(1)}, \dots , \mu^{((k)} \\ |\mu^{(i)}|=n  \\ \sum_{i=1}^k \ell^*(\mu^{(i)})=d }}
 m_\lambda({\bf c}) H(\mu^{(1)}, \dots , \mu^{(k)}, \mu, \nu)
\ee
with the sum  is over partitions $\lambda$ of length $k$,  and weight $d$ whose parts are $\{\ell^*(\mu^{(1)}), \dots, \ell^*(\mu^{(k)})\}$.

\bigskip
\noindent 
\small{ {\it Acknowledgements.}  The author would  like to thank G. Borot,  L.~Chekhov, M.~Guay-Paquet, 
M.~Kazarian,  S.~Lando and A.Yu.~Orlov for helpful discussions.}


\newcommand{\arxiv}[1]{\href{http://arxiv.org/abs/#1}{arXiv:{#1}}}

\bigskip
\noindent

\end{document}